\documentclass[AMA,STIX1COL]{WileyNJD-v2}

\usepackage{caption,tabularx,booktabs}
\newcommand{\Tau}{\mathrm{T}}
 
\usepackage{scalerel,stackengine}
\stackMath
\newcommand\reallywidehat[1]{%
\savestack{\tmpbox}{\stretchto{%
  \scaleto{%
    \scalerel*[\widthof{\ensuremath{#1}}]{\kern-.6pt\bigwedge\kern-.6pt}%
    {\rule[-\textheight/2]{1ex}{\textheight}}
  }{\textheight}%
}{0.5ex}}%
\stackon[1pt]{#1}{\tmpbox}%
}

\articletype{Research Article}%

\begin{document}

\title{Measurement Error in Meta-Analysis (MEMA) - \\ a Bayesian framework for continuous outcome data subject to non-differential measurement error}

\author[1]{Harlan Campbell*}

\author[2]{Valentijn M.T. {de Jong}}

\author[3]{Lauren Maxwell}

\author[3,5]{Thomas Jaenisch}

\author[2,4]{Thomas P.A. Debray}

\author[1]{Paul Gustafson}

\authormark{CAMPBELL \textsc{et al}}

\address[1]{\orgdiv{Department of Statistics}, \orgname{ University of British Columbia}, \orgaddress{\state{BC}, \country{Canada}}}

\address[2]{\orgdiv{Julius Center for Health Sciences and Primary Care}, \orgname{Utrecht University}, \orgaddress{\state{Utrecht}, \country{The Netherlands}}}

\address[3]{\orgdiv{Heidelberg Institute for Global Health}, \orgname{Heidelberg University Hospital}, \orgaddress{\state{Heidelberg}, \country{Germany}}}

\address[4]{\orgdiv{Cochrane Netherlands, Julius Center for Health Sciences and Primary Care}, \orgname{University Medical Center Utrecht, Utrecht University}, \orgaddress{\state{Utrecht}, \country{The Netherlands}}}

\address[5]{\orgdiv{Department of Epidemiology}, \orgname{Colorado School of Public Health}, \orgaddress{\state{CO}, \country{USA}}}

\corres{*Harlan Campbell,  \email{harlan.campbell@stat.ubc.ca}}


\abstract[Summary]{Ideally, a meta-analysis will summarize data from several unbiased studies. \textcolor{black}{ Here we look into the less than ideal situation in which contributing studies may be compromised by non-differential measurement error \textcolor{black}{ in the exposure variable.}  Specifically, we consider a meta-analysis for the association between a continuous outcome variable and one or more continuous exposure variables, where the associations may be quantified as regression coefficients of a linear regression model.}  A flexible Bayesian framework is developed which allows one to obtain appropriate point and interval estimates with varying degrees of prior knowledge about the magnitude of the measurement error.  We also demonstrate how, if individual-participant data (IPD) are available, the Bayesian meta-analysis model can adjust for multiple participant-level covariates, these being  measured with or without measurement error.}


\maketitle

\section{Introduction}\label{sec1}

Increasingly often, traditional meta-analysis methods are used to synthesize results from observational studies such as epidemiological surveys, cohort studies, and diagnostic test accuracy studies   \cite{debray2017guide, leeflang2008systematic, wells2013checklists}.  Observational studies are, by definition, non-randomized and are notoriously prone to a \textcolor{black}{wide range of biases, including selection bias and bias due to unobserved confounding \cite{hammer2009avoiding}.  One important bias that receives relatively little attention is measurement bias.  Since exposure variables} in an observational study are typically measured using imperfect tools (e.g., questionnaires, surveys, public health records), results are susceptible to ``bias caused by measurement error.''\cite{spiegelman1997regression}  \textcolor{black}{  Our focus will be on measurement error which we define as the error due to inaccurate measuring of the exposure variable(s).}

 If the measurement error affecting a particular study is of known magnitude, adjustment for measurement bias can be achieved by modifying the study's effect size estimate and uncertainty interval prior to its inclusion in a meta-analysis \cite{woodhouse1996adjusting}.  Typically, however, the magnitude of measurement error in any particular study is unknown, and appropriate adjustments are rarely done \cite{fosgate2006non, brakenhoff2018measurement}.  To be clear, issues of measurement error are not restricted to observational studies.  Indeed, measurement error is potentially problematic for a wide range of research studies regardless of study design. However, the assumption of no measurement error (or that measurement error does not affect the results) becomes more difficult to defend when the exposure of interest is not randomized or when variables of interest are difficult to quantify (e.g. no gold standard measurement tool exists, social constructs, stigmatized behaviors).

\textcolor{black}{
 In a meta-analysis of observational studies, failure to acknowledge and appropriately adjust for the possibility of measurement error amongst contributing studies will no doubt weaken or even invalidate the overall results \cite{thompson2010proposed}.  Yet measurement error has received relatively little attention in the meta-analysis literature.  
  Hunter and Schmidt (2004) discuss various pragmatic statistical approaches to correct for the impact of known measurement error \cite{hunter2004methods}  (and see more recently Wiernik et al. (2020) \cite{wiernik2020obtaining}).  While practical, these approaches fall short if the degree of measurement error is unknown.  Other work includes Carroll et al. (1991) \cite{carroll1991meta} who consider the merits of various attenuation factors to correct for measurement error in a meta-analysis.   These methods are developed ``[a]ssuming that data are available for consistent estimation of [the attenuation factors]''\cite{carroll1991meta}.  When such data are unavailable, the proposed attenuation factors fail to provide adequate adjustment.}
  
\textcolor{black}{In the applied literature, Zeisser et al. (2014) \cite{zeisser2014methodological} discuss the impact of likely exposure misclassification (i.e., measurement error of a binary exposure variable) in a meta-analysis estimating the relationship between alcohol consumption and breast cancer.  More recently, Lian et al. (2019) \cite{lian2019bayesian} introduce Bayesian meta-analysis models for binary outcomes accounting for exposure misclassification. These models are well designed but do not consider the possibility of measurement error in a continuous exposure or how to address continuous outcome data.}

 \textcolor{black}{Bayesian methods for handling measurement error are well established for single studies and offer ``a number of statistical advantages''  \cite{bartlett2018bayesian} due to their inherent  flexibility to handle more complicated data structures.} Bayesian methods also offer a ``number of specific advantages'' for meta-analysis; see Sutton and Abrams (2001) \cite{sutton2001bayesian}.  For instance, unlike Bayesian models, frequentist meta-analysis models are known to have difficulty estimating variance parameters if these parameters are near-zero, particularly when sample sizes are small \cite{mcneish2016using, chung2013avoiding}.  Also, Bayesian models also offer substantial flexibility for handling complicated data structures that may arise with multiple covariates and the possibility of measurement error \cite{hossain2009bayesian}.  (That being said, frequentist models are often easily implemented with standard statistical packages whereas Bayesian models may require a certain amount of customization \cite{rover2017bayesian}).
 
  In this paper, we consider a meta-analysis of observational studies with continuous outcome and exposure variables in which (a subset of) contributing studies may be compromised by a potentially unknown degree of non-differential measurement error, i.e., error in the exposure variable(s) that is conditionally independent of the outcome variable \cite{brenner1994varied}.   We then develop a Bayesian hierarchical model to adjust for the measurement error when either aggregate study-level data or individual participant-level data (IPD) are available.  In Section 2, we outline a proposed Bayesian framework for the case of meta-analysis with measurement error in a single exposure variable and in Section 3, we generalize this framework for the case of multiple explanatory variables.  We conclude with a summary of findings in Section 4.

\section{Meta-analysis of simple linear regression aggregate data}
\label{sec:linmodel}
\subsection{A traditional random-effects meta-analysis}
\label{sec:simplewithout}
\textcolor{black}{
Suppose we have data from $K$ independent observational studies for a meta-analysis and, for each of these studies, the exposure and the outcome are continuous variables.  We begin by describing some basic distributional assumptions for the underlying data.  }
 
Let \(\left(X_{j}^{[k]}, Y_{j}^{[k]}\right)\) be the exposure and outcome for the \(j\)-th observation in the \(k\)-th  study.  We assume that the exposure and outcome are related by means of a conditional Normal distribution such that: 
\(Y_{j}^{[k]} | X_{j}^{[k]} \sim  \mathcal{N}\left(\alpha^{[k]}+\beta^{[k]} X_{j}^{[k]}, \sigma^{[k]2}\right)\); for $k = 1,\ldots,K$ and $j$ in $1,\ldots,n^{[k]}$.  Furthermore, we assume that each study has its own exposure distribution governed by a Normal distribution: \(X_{j}^{[k]} \sim \mathcal{N}\left(\mu^{[k]}, \lambda^{[k]2}\right)\).  Finally, the study specific parameters, ${\alpha}^{[k]}$ and ${\beta}^{[k]}$, are related to one another such that:
 
\begin{align} 
\begin{array}{c}
\alpha^{[k]}\\
\beta^{[k]}
\end{array}
\sim \mathcal{N}\left(\begin{pmatrix}
\xi \\
\theta
\end{pmatrix}, \quad \begin{pmatrix}  
\omega ^{2}& \rho\omega\tau \\
\rho\omega\tau  & \tau^{2}
\end{pmatrix}\right).
\label{eq:multinorm_alphabeta_true}
\end{align}

\noindent for $k = 1,\ldots,K$, where $\xi$ is the overall mean intercept parameter, $\omega ^{2}$ represents the variance in intercepts across studies, $\theta$ is the overall mean slope parameter (and the main ``parameter of interest''), $\tau^{2}$ is the variance of slopes across studies, and $\rho$ is the correlation of the regression coefficients.



For each of the $K$ studies, a standard simple linear regression model could be fit to the outcome and exposure data to obtain least-squares parameter estimates, $\hat{\alpha}^{[k]}$ and $\hat{\beta}^{[k]}$, which will follow, according to standard theory \citep{myers1990classical}, a bivariate Normal distribution such that,  for $k = 1,\ldots,K$: 

\begin{align} 
\begin{array}{c|c}
\hat{\alpha}^{[k]} & \alpha^{[k]}\\
\hat{\beta}^{[k]} & \beta^{[k]}
\end{array}
\sim \mathcal{N}\left(\begin{pmatrix}
\alpha^{[k]} \\
\beta^{[k]}
\end{pmatrix}, \quad \begin{pmatrix}  
\Sigma_{11}^{[k]} & \Sigma_{12}^{[k]} \\
\Sigma_{12}^{[k]}  & \Sigma_{22}^{[k]} 
\end{pmatrix}\right),
\label{eq:multinorm_alphabeta}
\end{align}
%
%
%
\begin{equation}
\textrm{where:  } \quad \Sigma_{11}^{[k]} =   \frac{(\lambda^{[k]2}+\mu^{[k]2})\times\sigma^{[k]2}}{\lambda^{[k]2}\times n^{[k]}} ; \quad  \textrm{} 
\label{eq:var_alpha}
\end{equation}
\begin{equation}
\Sigma_{12}^{[k]} = - \mu^{[k]} \times \frac{\sigma^{[k]2}}{\lambda^{[k]2}\times n^{[k]}};
    \label{eq:covar_alphabeta}
\end{equation}
\begin{equation}
 \quad \textrm{ and} \quad \Sigma_{22}^{[k]} =  \frac{\sigma^{[k]2}}{\lambda^{[k]2}\times n^{[k]}}.
\label{eq:var_beta}
\end{equation}

A meta-analysis will typically combine the summary statistics reported in each contributing study to obtain an overall estimate for the parameter(s) of interest.  If the value of $\hat{\beta}^{[k]}$ and its standard error, $\textrm{se}(\hat{\beta}^{[k]})$, are available for $k = 1,\ldots,K$, the primary parameter of interest, $\theta$, can be estimated in a standard univariate random-effects meta-analysis \citep{brockwell2001comparison} in which:

\textcolor{black}{
\begin{equation}
\hat{\beta}^{[k]} \sim \mathcal{N}(\theta, (\textrm{se}(\hat{\beta}^{[k]}))^{2} + \tau^{2}).
\label{eq:MA_justbeta}
\end{equation}
}

\noindent Or if, in an admittedly rare situation (Becker et al. (2007) \cite{becker2007synthesis} point to Crouch (1995,1996)\cite{crouch1995meta, crouch1996demand} and Lau et al. (1999) \cite{lau1999effects} as examples), data are also available for $\hat{\alpha}^{[k]}$, $\textrm{se}(\hat{\alpha}^{[k]})$ and $\hat{\sigma}^{[k]2}$ for $k = 1,\ldots,K$, one may fit a bivariate meta-analysis model \citep{novick1972estimating, schmid2020handbook} in which:
\begin{align} 
\begin{pmatrix}
\hat{\alpha}^{[k]}\\
\hat{\beta}^{[k]}
\end{pmatrix}
\sim \mathcal{N}\left(\begin{pmatrix}
\xi \\
\theta
\end{pmatrix}, \quad \begin{pmatrix}  
\Sigma_{11}^{[k]} +  \omega ^{2} & \Sigma_{12}^{[k]} +\rho\omega\tau \\
\Sigma_{12}^{[k]} + \rho\omega\tau& \Sigma_{22}^{[k]} +  \tau^{2}
\end{pmatrix}\right),
\label{eq:multinorm_alphabeta_uncond}
\end{align}

\noindent where $\Sigma_{11}^{[k]}$, $\Sigma_{12}^{[k]}$, and $\Sigma_{22}^{[k]}$ are given by equations (\ref{eq:var_alpha}), (\ref{eq:covar_alphabeta}), and (\ref{eq:var_beta}) with the $\mu^{[k]}$, $\sigma^{[k]}$, and $\lambda^{[k]}$ parameters assumed to be known (i.e., measured without error) and equal to:

\begin{equation}
{\sigma}^{[k]} = \hat{\sigma}^{[k]} , \quad \quad {\lambda}^{[k]} = \frac{\hat{\sigma}^{[k]}}{\textrm{se}(\hat{\beta}^{[k]})\sqrt{(n^{[k]}-1) }} , 
\end{equation}
\noindent and
\begin{equation}
\quad {\mu}^{[k]} = \sqrt{\Big(\frac{\textrm{se}(\hat{\alpha}^{[k]})}{\textrm{se}(\hat{\beta}^{[k]})}\Big)^{2} - \frac{\hat{\sigma}^{[k]2}}{n^{[k]}\times(\textrm{se}(\hat{\beta}^{[k]}))^{2}}  }.
\end{equation}

\vspace{0.2cm}
\noindent  \textcolor{black}{Note that with a sufficient amount of data, the uncertainty surrounding the  $\mu^{[k]}$, $\sigma^{[k]}$, and $\lambda^{[k]}$ parameters will be quite small \cite{sutton2001bayesian}, and the simplifying assumption that the $\mu^{[k]}$, $\sigma^{[k]}$, and $\lambda^{[k]}$ parameters are known should make little practical difference \cite{chung2013avoiding}.  However, if one wished to properly account the additional uncertainty of $\mu^{[k]}$, $\sigma^{[k]}$, and $\lambda^{[k]}$, a suitable strategy would be to create pseudo-values for \(\left(X_{j}^{[k]}, Y_{j}^{[k]}\right)\) , for $k = 1,\ldots,K$ and $j$ in $1,\ldots,n^{[k]}$, using the observed sufficient statistics (if these were all available).  This pseudo individual participant level data would have the same likelihood as the true unknown underlying IPD and could then be fit --as if it were the true data-- with an IPD-meta-analysis model (as in Section \ref{sec:multiwme}); see Papadimitropoulou et al. (2020)  \cite{papadimitropoulou2020meta}.}

\textcolor{black}{
Inference for either the univariate or bivariate meta-analysis  (i.e., for either (\ref{eq:MA_justbeta}) or (\ref{eq:multinorm_alphabeta_uncond})) can be done within either a frequentist or a Bayesian framework \citep{becker2007synthesis, kim2019meta,williams2018bayesian}.   However, as discussed in the introduction, there are several reasons why a Bayesian approach may be advantageous.}  A Bayesian model requires defining priors for all of the unknown parameters and, for better or worse, the performance of any Bayesian estimator will depend on the choice of these priors.   Particularly when few data are available, the choice of priors can substantially influence the posterior \cite{lambert2005vague, berger2013statistical, burke2018bayesian}.    In the examples considered throughout this paper, our strategy will be to adopt wide Normal distributions (with variance of 100) for the mean parameters, weakly-informative half-Cauchy priors (with scale parameter of 2) for the variance parameters, and a uniform distribution for the correlation parameter; following the recommendations of Polson et al. (2012) \cite{polson2012half} and the simulation results of Williams et al. (2018) \cite{williams2018bayesian}.  

Before going on to discuss measurement error, let us briefly demonstrate how standard Bayesian univariate and bivariate meta-analysis models (\textit{BayesMA}) can be used in an analysis of some simple illustrative data.

\subsubsection{Example: the NELS88 dataset}
\label{sec:nels88_1}

The NELS88 dataset has been used previously as an example dataset by Becker et al. (2007) \cite{becker2007synthesis} and is from a survey of U.S. grade 10 high-school students in 1988 from over 1,000 schools.   Becker et al. (2007) \cite{becker2007synthesis} include for analysis only the 13 schools with samples of a minimum of 45 students ($n^{[k]} \ge 45; k =1,\ldots,13$) and consider each school as an independent study for meta-analysis.  We will use the same subset of schools for our example analysis.  The outcome of interest, $Y_{j}^{[k]}$, will be the science achievement test score, and the exposure of interest, $X_{j}^{[k]}$, will be the reading test score, for the $j$-th student in the $k$-th school.  The total sample size is $N = \sum_{k=1}^{K}{n^{[k]}}=664$ students from $K=13$ different schools.  
  
  Table \ref{tab:nels88} displays the aggregate data from the NELS88 dataset required for both the univariate and bivariate meta-analyses.  We fit these data with \textit{BayesMA} models defined in Section \ref{sec:simplewithout} with the following priors:
  
\hspace{1cm} $ \theta \sim \mathcal{N}(0, 100);  $ (mean of 0, variance of 100) 

\hspace{1cm} $  \xi  \sim \mathcal{N}(0, 100); $ 

\hspace{1cm} $  \tau \sim \textrm{half-Cauchy}(0, 2), \quad \tau > 0 $; \hspace{0.5cm} (location of 0, scale of 2); 

\hspace{1cm}   $  \omega  \sim \textrm{half-Cauchy}(0, 2), \quad  \omega >0$; and

\hspace{1cm}   $  \rho  \sim \textrm{Uniform}(-1, 1)$.

 All models in this paper are fit using the probabilistic programming language JAGS which employs the Gibbs sampling Markov chain Monte Carlo (MCMC) algorithm and is compatible with the R statistical programming language \cite{kruschke2014doing}.  Each model is fit based on 100,000 Monte Carlo draws from each of three chains (thinning of 10), and for each we report posterior medians and equal-tailed 95\% credible intervals.  \textcolor{black}{ Note that the \textit{BayesMA} model (or something similar) could be easily fit in a frequentist manner with standard statistical packages.  However, this is not the case for the \textit{BMEMA} model we introduce in Section \ref{sec:univariate_withME}.} 
  
\begin{figure}[h]
      \centering
      \includegraphics[width=17cm]{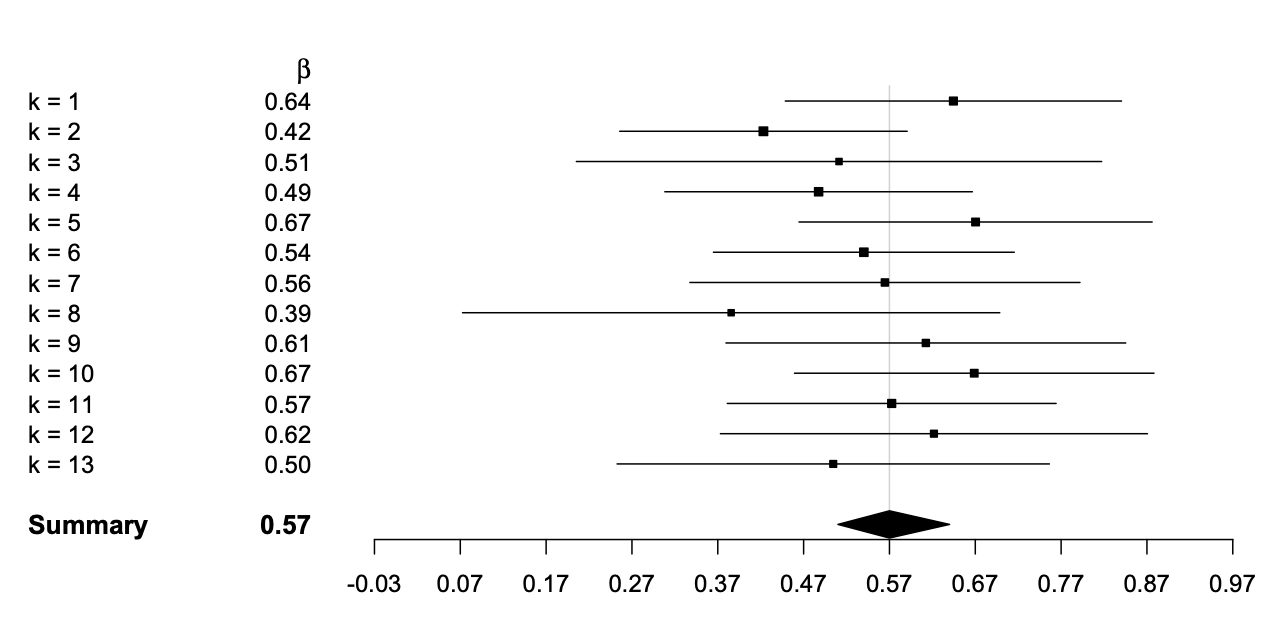}
      \caption{Forest plot for the meta-analysis of the NELS88 aggregate data (Table \ref{tab:nels88}). The $K=13$ black squares correspond to the $\hat{\beta}^{[k]}$ values, for $k = 1,\ldots,K$;  with horizontal lines corresponding to symmetrical $1.96\times\textrm{se}(\hat{\beta}^{[k]})$ confidence intervals.  The \textit{ BayesMA} posterior estimate of $\theta$ is plotted as a diamond (labelled ``Summary''), the lateral points of which indicate the equal-tailed 95\% credible interval (i.e., the 2.5\% and 97.5\% quantiles) for this estimate.}
      \label{fig:forest_plot1}
  \end{figure}

Results for the univariate and bivariate models are very similar but not identical.  For the univariate model, we obtain posterior medians: $\hat{\theta}=0.56$ with 95\% equal-tailed credible interval of ${CI(\theta)}_{95\%} = [0.49, 0.62]$; and $\hat{\tau} =   0.04$.  For the bivariate model, we obtain the posterior medians: $\hat{\theta}=0.57$ with 95\% equal-tailed credible interval of ${CI(\theta)}_{95\%} = [0.51, 0.64]$;  $\hat{\tau} =   0.04$; $\reallywidehat { \xi}   = 5.40$ with 95\% equal-tailed credible interval of  ${CI(\xi)}_{95\%} = [4.05, 6.86]$;  $\reallywidehat{\omega} = 1.85$; and $\hat{\rho}=0.06$.  Figure \ref{fig:forest_plot1} displays a forest plot summarizing the primary analysis results for the bivariate model.

\begin{table}[p]
\centering
\begin{tabular}{rc|cc|rrrrr}
\multicolumn{8}{l}{\textbf{The NELS88 aggregate data} } &\\   
 & sample  & \multicolumn{2}{l|}{data required for}  &  \multicolumn{4}{l}{data required for}   \\ 
 &  size & \multicolumn{2}{l|}{univariate model}  &  \multicolumn{4}{l}{bivariate model}   \\ \hline
   $k$ & $n^{[k]}$  & $\hat{\beta}^{[k]}$ & $\textrm{se}(\hat{\beta}^{[k]})$ & $\hat{\alpha}^{[k]}$ & $\textrm{se}(\hat{\alpha}^{[k]})$  & $\hat{\beta}^{[k]}$ & $\textrm{se}(\hat{\beta}^{[k]})$ & $\hat{\sigma}^{[k]2}$\\
  \hline
1 & 45 & 0.64 & 0.10 & 4.99 & 1.53 & 0.64 & 0.10 & 12.28 \\ 
  2 & 64 & 0.42 & 0.09 & 7.91 & 1.16 & 0.42 & 0.09 & 19.50 \\ 
  3 & 47 & 0.51 & 0.16 & 8.46 & 2.71 & 0.51 & 0.16 & 11.48 \\ 
  4 & 45 & 0.49 & 0.09 & 2.95 & 1.19 & 0.49 & 0.09 & 15.71 \\ 
  5 & 45 & 0.67 & 0.10 & 1.75 & 1.07 & 0.67 & 0.10 & 15.15 \\ 
  6 & 59 & 0.54 & 0.09 & 5.53 & 1.24 & 0.54 & 0.09 & 19.45 \\ 
  7 & 56 & 0.56 & 0.12 & 5.48 & 1.77 & 0.56 & 0.12 & 17.31 \\ 
  8 & 45 & 0.39 & 0.16 & 9.69 & 2.78 & 0.39 & 0.16 & 13.51 \\ 
  9 & 51 & 0.61 & 0.12 & 6.60 & 2.15 & 0.61 & 0.12 & 6.93 \\ 
  10 & 67 & 0.67 & 0.11 & 6.33 & 1.84 & 0.67 & 0.11 & 13.88 \\ 
  11 & 48 & 0.57 & 0.10 & 4.03 & 1.23 & 0.57 & 0.10 & 11.62 \\ 
  12 & 45 & 0.62 & 0.13 & 6.39 & 1.85 & 0.62 & 0.13 & 20.55 \\ 
  13 & 47 & 0.50 & 0.13 & 4.02 & 1.85 & 0.50 & 0.13 & 15.16 \\ 
   \hline
\end{tabular}
\caption{The NELS88 data- Aggregate data required for the univariate and bivariate meta-analyses in Section \ref{sec:nels88_1}, as obtained from the NELS88 dataset.}
\label{tab:nels88}
\end{table}

\subsection{Adjusting for non-differential measurement error}
\label{sec:univariate_withME}

Now suppose that each study is hampered by a certain amount of classical and non-differential measurement error.  The assumption of \textit{non-differential} measurement error refers to \textcolor{black}{the assumption that the} distribution of the surrogate exposures, $X^{*}$, depends only on the actual exposure variables, $X$, and not on the response variable or other variables in the model.   In other words, we assume that the conditional distribution of $(X^{*}|X,Y)$ is identical to the conditional distribution of $(X^{*}|X)$. 

In this situation, we wish to determine the relationship between the outcome, $Y$, and the exposure, $X$, with data based instead on measuring $Y$ and $X^{*}$.  Ignoring the measurement error in such a situation will bias the estimates of the regression slope coefficients towards the null  (i.e., will bias $\hat{\beta}$ towards 0); see Hutcheon et al. (2010) \cite{hutcheon2010random} for an excellent review.  We assume the vector of independent surrogates, $X^{*}$, arises from a classical additive measurement error model:
\begin{equation}
X_{j}^{[k]*}|X_{j}^{[k]} \sim  \mathcal{N}\left(X_{j}^{[k]} , \phi^{[k]2}\right)  \textrm{, for $j$ in $1,\ldots,n^{[k]}$, }
\label{eq:nondiff}
\end{equation}
 \noindent  for $k = 1,\ldots,K$, where $\phi^{2[k]}$ corresponds to the variance in measurement error for the $k$-th study.  To be clear, for each study, the average measurement error is zero but individual measurements can be biased, such that:
 
 \begin{equation}
X_{j}^{[k]*} \sim \mathcal{N}\left(\mu^{[k]}, \lambda^{[k]*2}\right)  
\label{eq:nondiff2}
\end{equation}

\noindent where $\lambda^{[k]*2}=\lambda^{[k]2} + \phi^{[k]2}$; for $j$ in $1,\ldots,n^{[k]}$ and  for $k = 1,\ldots,K$.

Let \(\gamma^{[k]}= \lambda^{[k]2}/(\lambda^{[k]2} + \phi^{[k]2}) = \left(1+\phi^{[k]2} / \lambda^{[k]2}\right)^{-1} < 1 \) 
be the ``attenuation factor'' for the \(k\)-th study, for $k = 1,\ldots,K$.  The range of values for $\phi^{[k]}$ is therefore restricted to: $0 \le \phi^{[k]} \le  \lambda^{[k]*}$.  Linear regression coefficients estimated for each of the individual studies will be biased and governed by:
\begin{align} 
\begin{array}{c|c}
\hat{\alpha}^{[k]*} & \alpha^{[k]}\\
\hat{\beta}^{[k]*} & \beta^{[k]}
\end{array}
\sim \mathcal{MVN}\left(\begin{pmatrix}
\alpha^{[k]} + (1-\gamma^{[k]}) \beta^{[k]} \mu^{[k]} \\
 \gamma^{[k]} {\beta^{[k]}} 
\end{pmatrix}, \quad  \begin{pmatrix}  
\Sigma_{11}^{[k]*} & \Sigma_{12}^{[k]*} \\
\Sigma_{12}^{[k]*} & \Sigma_{22}^{[k]*}
\end{pmatrix}\right),
\label{eq:multinorm_alphabeta_star}
\end{align}
\begin{equation}
\textrm{where:  } \quad \Sigma_{11}^{[k]*}  =   \frac{(\lambda^{[k]*2}+\mu^{[k]*2})\times\sigma^{[k]*2}}{\lambda^{[k]*2}\times n^{[k]}} ; \quad  \textrm{} 
  \label{eq:sigma11_me}
\end{equation}
\begin{equation}
\quad \Sigma_{12}^{[k]*} = - \mu^{[k]*} \times \frac{\sigma^{[k]*2}}{\lambda^{[k]*2}\times n^{[k]}};  
  \label{eq:sigma12_me}
\end{equation}
\begin{equation}
\quad \textrm{and } \Sigma_{22}^{[k]*}   =  \frac{\sigma^{[k]*2}}{\lambda^{[k]*2}\times n^{[k]}}; 
  \label{eq:sigma22_me}
\end{equation}
where $\mu^{[k]*} = \mu^{[k]}$; $\sigma^{[k]*2} =  \sigma^{[k]2} + (1-\gamma^{[k]}) (\beta^{[k]2})(\lambda^{[k]2})$; and 
$\lambda^{[k]*2} = \lambda^{[k]2} + \phi^{[k]2}$, for $k = 1,\ldots,K$.

 If we ignore measurement error, then, for the parameter of interest $\theta$, we will mistakenly target:
 \begin{equation}
     \theta^{*}=\textrm{E}\left(\gamma^{[k]} {\beta^{[k]}}\right) =\textrm{E}\left\{\left(1+  \frac{{\phi^{[k]}}^{2}} { {\lambda^{[k]}}^{2}}\right)^{-1} \times  {\beta^{[k]}}\right\},
 \end{equation} 
 \noindent in place of \(\theta=\textrm{E}\left({\beta^{[k]}}\right) \), for all $k = 1,\ldots,K$.  As such, if we presume independence of \(\beta^{[k]}\) and \(\left(\phi^{[k]}, \lambda^{[k]}\right)\), then we have:
 \begin{equation}
 \theta^{*}=\textrm{E}\left({\gamma^{[k]}}\right) \theta .
  \label{eq:theta_me}
 \end{equation}

 This is intuitive: the attenuation factor induced by measurement error in estimating the typical exposure-outcome association is the expectation of the study-specific attenuation factors.  This suggests that an unbiased estimate of the overall effect, $\theta$, can be derived analytically  if one knows the cross-study average degree of measurement error.  In other words, it is not necessary to know each individual value of $\phi^{[k]}$, for $k = 1,\ldots,K$.  One need only know where the distribution of the $\phi^{[k]}$s is centered in order to adequately adjust the meta-analytic point estimate of $\theta$ in the presence of measurement error.

The bias in estimating \(\tau^{2}\) is perhaps less intuitive. If we ignore the presence of measurement error, then our estimation procedures will mistakenly target: $\tau^{*2}=\operatorname{Var}\left({\beta}^{[k]*}\right)=\operatorname{Var}\left({\gamma^{[k]}} {\beta^{[k]}}\right) $, for all $k = 1,\ldots,K$.   Invoking independence of $\gamma^{[k]}$ and $\beta^{[k]}$, we have:
\begin{equation}
\tau^{*2}=\left\{\textrm{E}\left({\gamma^{[k]}}\right)\right\}^{2} \tau^{2}+\operatorname{Var}\left({\gamma^{[k]}}\right)\left(\tau^{2}+\theta^{2}\right),
\label{eq:tau_me}
\end{equation}

\noindent or alternatively: $\tau^{* 2}=\textrm{E}\left({\gamma^{[k]}}^{2}\right) \tau^{2}+\operatorname{Var}\left({\gamma^{[k]}}\right) \theta^{2}$.  The first term alone speaks to underestimating cross-study heterogeneity if we ignore measurement error, since we know $0 < \textrm{E}\left({\gamma^{[k]}}^{2}\right)<1$, for all $k = 1,\ldots, K$.  However, the second term could counteract this. More precisely, if the magnitude of the cross-study variation in measurement error (i.e., $\operatorname{Var}({\gamma^{[k]}})$) and/or the average effect size (i.e., $\theta$) are large enough, we could end up overestimating cross-study heterogeneity instead.  It follows that, without any cross-study heterogeneity, i.e., when $\tau^{2}=0$, equation (\ref{eq:tau_me}) reduces to $\tau^{*2}=\operatorname{Var}\left({\gamma^{[k]}}\right)\theta^{2}$.  Therefore, if the cross-study variability in measurement error is sufficiently large, one could erroneously select a random-effects model ($\tau^{2}>0$) instead of the correct fixed-effects model  ($\tau^{2}=0$).

For the bivariate model, the bias in estimating $ \xi $ and $ \omega ^{2}$ must also be considered.  If we ignore measurement error, then our estimation procedures will mistakenly target, for $k = 1,\ldots,K$: 

\begin{equation}
 \xi ^{*} = \textrm{E}\{\alpha^{[k]} + (1-\gamma^{[k]}) \beta^{[k]} \mu^{[k]}\} =  \xi  + \textrm{E}\{(1-\gamma^{[k]}) \beta^{[k]} \mu^{[k]}\},
 \label{eq:xi_me}
\end{equation}
\noindent and
\begin{equation}
\omega ^{*2}=\operatorname{Var}\left({\alpha}^{[k]*}\right)=\operatorname{Var}\left(\alpha^{[k]} + (1-\gamma^{[k]}) \beta^{[k]} \mu^{[k]}\right).
 \label{eq:omega_me}
\end{equation}

 The correlation between different elements of the data will also be impacted by measurement error.  For example, in the absence of measurement error (i.e., when $\gamma^{[k]} = 1$), the estimators, $\hat{\beta}^{*[k]}$ and $\hat{\lambda}^{*[k]}$ will be entirely independent.  However, in the presence of heterogeneous measurement error, this is no longer the case.  A large amount of measurement error will lead to a larger value of $\hat{\lambda}^{*[k]}$ and simultaneously to a smaller value of $\hat{\beta}^{*[k]}$.  As such, if the $k$-th study has a relatively (as compared to the other studies) small value of $\hat{\beta}^{*[k]}$ and a relatively large value of $\hat{\lambda}^{*[k]}$, this suggests that (relative to other studies) it may be compromised by a substantial degree of measurement error.  
 
 The standard \textit{BayesMA} models described in Section \ref{sec:simplewithout}, can be adapted to account for non-differential measurement error in a relatively straightforward manner.  The univariate \textit{BMEMA} (Bayesian model for Measurement Error in Meta-Analysis) can be defined in two parts as:

\textcolor{black}{
\begin{equation}
\hat{\beta}^{[k]*}|\beta^{[k]} \sim \mathcal{N}(\gamma^{[k]}\beta^{[k]}, (\textrm{se}(\hat{\beta}^{[k]}))^{*2}) \quad \quad   \textrm{and} \quad \quad \beta^{[k]} \sim \mathcal{N}(\theta, \tau^{2}),
\end{equation}
}
\textcolor{black}{\noindent where $\textrm{se}(\hat{\beta}^{[k]})$ is the standard error for $\hat{\beta}$ as reported by a study potentially biased due to measurement error; and where, if data for ${\lambda}^{[k]*2}$ are available, we can define:}
\textcolor{black}{
\begin{equation}
\gamma^{[k]} = \Big(1 + \frac{\phi^{[k]2}}{({\lambda}^{[k]*2} - \phi^{[k]2})} \Big)^{-1}.
\end{equation}
}

 The bivariate \textit{BMEMA} model can also defined in two-parts with the conditional bivariate normal likelihood for $(\hat{\alpha}^{[k]*},  \hat{\beta}^{[k]*} | \alpha^{[k]}, \beta^{[k]})$ as specified by equation (\ref{eq:multinorm_alphabeta_star}), and the bivariate normal likelihood for $\alpha^{[k]}$ and $\beta^{[k]}$ as specified by equation (\ref{eq:multinorm_alphabeta_true}).

 Knowledge about the magnitude of measurement error in a study may come from a variety of sources, e.g. replicate measurements, validation data, or expert opinion.  In some scenarios, one might have a subset of studies for which $\gamma^{[k]}$ is known and equal to 1 (i.e., have data from some studies known to be unaffected by measurement error).  We will focus on how one might proceed under such a scenario.  Without loss of generality, suppose a subset of ``gold standard'' studies is the first $k^{'}$ studies, such that for $k = 1, \ldots, k^{'}$, we have $\phi^{[k]2}=0$ and $\gamma^{[k]}=1$.  Thus, in a situation where all studies are known to be unbiased by measurement error, $k^{'}$ will equal $K$ and the \textit{ BayesMA} and \textit{ BMEMA} models will be identical.  

For a scenario in which $0 < k^{'} < K$, our strategy will depend on whether or not, for $k =  k^{'}+1,\ldots,K$, data are available for $\hat{\lambda}^{[k]*}$, as this can serve as an upper bound on $\phi^{[k]}$ (recall that: $0 \le \phi^{[k]} \le  \lambda^{[k]*}$).  If  $\hat{\lambda}^{[k]*}$ data is not available, we can  simply specify a uniform prior on $\gamma^{[k]}$, such that, for $k =  (k^{'}+1),\ldots,K$:
 \begin{eqnarray}
 \gamma^{[k]} \sim& \textrm{Uniform}(0,1).
 \label{eq:uniform}
 \end{eqnarray}

\noindent Alternatively, if $\hat{\lambda}^{[k]*}$ is available for $k =  (k^{'} +1),\ldots,K$, we place an inverse-gamma prior on the study-specific $\phi^{[k]2}$ parameters, such that:
 \begin{eqnarray}
\phi^{[k]2}  \sim& \textrm{Inv-Gamma}(\zeta_{1}, \zeta_{2}); \quad \quad \textrm{and} \quad \zeta_{1} \sim \textrm{Exp}(\delta), \quad \zeta_{2} \sim \textrm{Exp}(\delta),
 \label{eq:invgamma}
 \end{eqnarray}
\noindent for $k = (k^{'}+1),\ldots,K$; and for  $\zeta_{1}>0$ and $\zeta_{2}>0$. Note that the mean and variance of the inverse-gamma distribution have that, \textit{ a priori}, $\textrm{E}(\phi^{[k]2} ) = \zeta_{2}/(\zeta_{1}-1)$ and $\textrm{Var}(\phi^{[k]2} ) = \zeta_{2}^{2}/((\zeta_{1} - 1)^{2}(\zeta_{1} - 2))$.  However, only values of $\phi^{[k]2} \le \lambda^{[k]*}$ will be consistent with the data.   In order to reflect very vague prior knowledge, we set $\delta=0.1$.

For a scenario in which $k^{'}=0$, things are not so straightforward.  Without any ``gold standard'' studies and without any specific knowledge regarding the degree of measurement error for each of the $K$ contributing studies, the parameters of interest may not be identifiable.   With this in mind, before moving on, we briefly consider an asymptotic argument for so-called ``partial identifiability'' by considering the degree to which $\theta$ can be estimated in the presence of unspecified measurement error. 

\subsubsection{Issues of identifiability}
\label{sec:unident}

Our logic is based on similar arguments for partial identifiability considered in Campbell et al. (2020) \cite{campbell2020bayesian}.  Presume that \textit{ a priori} defensible information about the amount of bias caused by measurement error in the $k$-th study is expressed in the form: $\gamma^{[k]}  \in [\underbar{$\gamma$}^{[k]}, 1]$, where $\underbar{$\gamma$}^{[k]}$ is an investigator-specified lower bound for the $k$-th attenuation factor.  Then the set of possible values for $\beta^{[k]}$, given $\beta^{[k]*}$, is restricted to:
 \begin{equation}  \label{eq:IFR(phi2)}
I_{k}(\underbar{$\gamma$}^{[k]}) = \Big[\frac{ \beta^{*[k]}}{\underbar{$\gamma$}^{[k]}} , \beta^{*[k]}\Big].
\end{equation}
To be clear, this represents the study-specific {identification interval} for $\beta^{[k]}$.
As $n^{[k]} \rightarrow \infty$, all values inside the interval remain plausible, while all values outside are ruled out \cite{manski2003partial}.  This is the essence of the \textit{ partial identification} inherent to this problem.

\begin{figure}
    \centering
    \includegraphics[width=14cm]{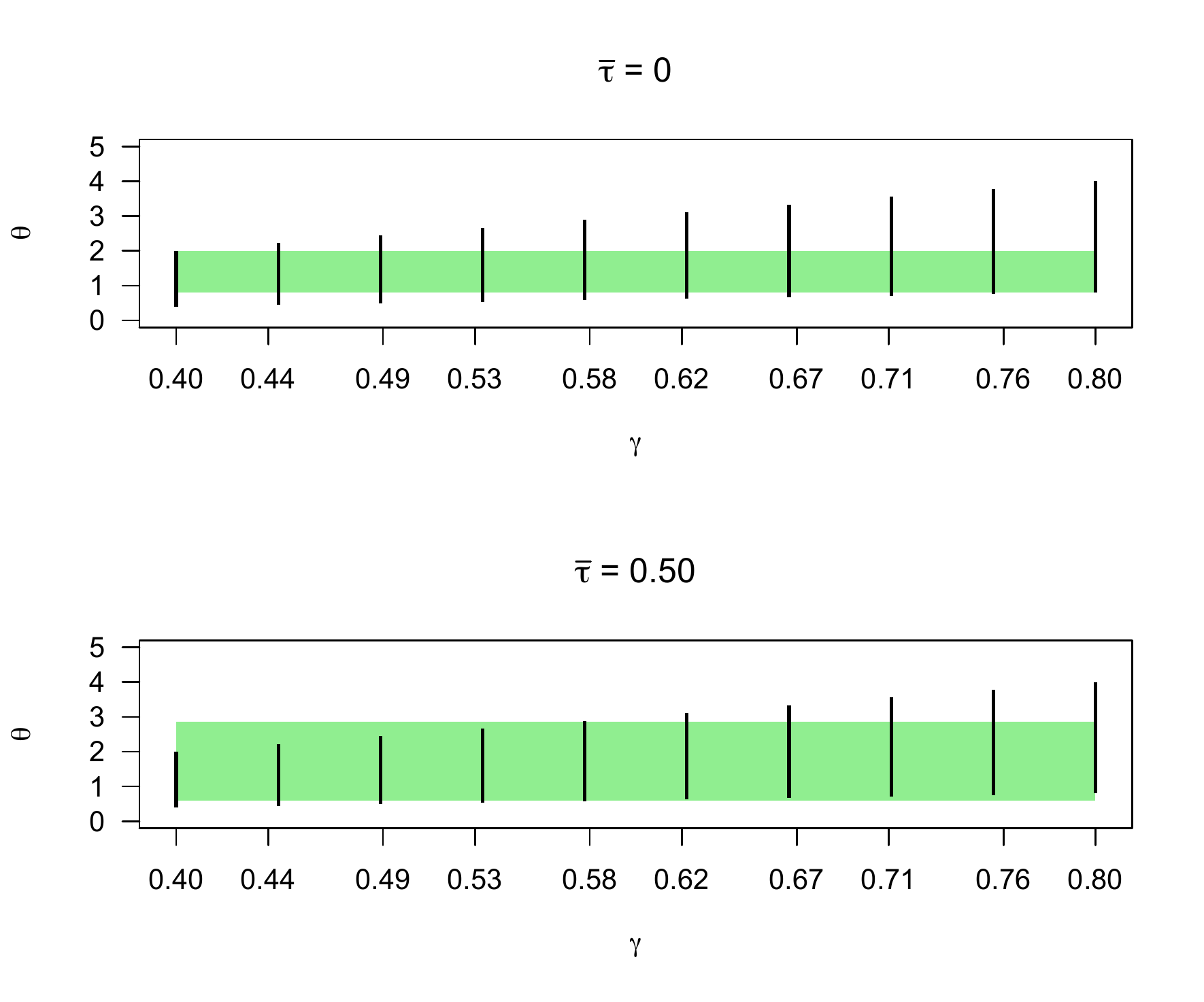}
    \caption{Black vertical lines correspond to study-specific identification intervals (i.e., the intervals for $\theta$ that are compatible with data from the corresponding study) and the green rectangle corresponds to the global identification interval.  The upper panel corresponds to assumption of $\bar{\tau}=0$ such that the global identification interval is simply the intersection of the individual study-specific identification intervals. The lower panel corresponds to  $\bar{\tau}$ = 0.50.}
    \label{fig:greenbox}
\end{figure}

Thinking now about the meta-analytic task of combining information, the $\gamma^{[k]}$ could exhibit considerable variation across studies while $\tau$ (i.e., the variation in $\beta^{[k]}$) could be small.  Suppose that $\tau$  does not exceed an investigator-specified upper bound of $\bar{\tau}$,  i.e.,  $\tau \leq \bar{\tau}$.  Then an identification region for $\theta$ can be specified as:
\begin{eqnarray} \label{eq-idef1}
I(\underbar{$\gamma$}, \bar{\tau})  =
\left\{ \theta : 
\tau \leq \bar{\tau}, 
\beta^{[k]} \in I_{k}(\underbar{$\gamma$}^{[k]}), \forall k \in \{1, \ldots, K\}
\right\}.
\label{eq:partialid}
\end{eqnarray}
Again, the interpretation is direct:
in the asymptotic limit, all values of $\theta$ inside this interval are compatible with the observed data, and all values outside are not.
The primary question of interest is whether this interval is narrow or wide under realistic scenarios, 
since this governs the extent to which we can learn about $\theta$ from the data.

In general, evaluating (\ref{eq-idef1}) for given inputs is an exercise in quadratic programming nested within a grid search, hence can be handled with standard numerical optimisation.   
%
%
However, the special ``fixed-effects''  case of $\bar{\tau}=0$ is noteworthy.
%
%
Mathematically, the case is much simpler, with (\ref{eq-idef1}) reducing to $I(\underbar{$\gamma$}, 0)  = \cap_{k}
I_{k}(\underbar{$\gamma$}^{[k]}).$  As intuition must have it, without heterogeneity, a putative value for $\theta$ is compatible with the observed data if and only if it is compatible with the data from \textit{ every} individual study.

To illustrate,  consider a scenario with $K=10$ studies, with $\theta=1.0$, and $\tau=0$ i.e., $\beta^{[k]}=1.0$, for $k = 1,\ldots,10$.  Suppose the observed $\hat{\beta}^{[k]*}$ values for these studies lie equally spaced between 0.40 and 0.80, (since the unknown $\gamma^{[k]}$ values range between $0.40$ and $0.80$).  Furthermore, say the investigator pre-specifies $\underbar{$\gamma$}^{[k]} = 0.2$ for all $k$.  The resulting study-specific identification intervals, $I_k$, are depicted by the black vertical lines in the upper panel of Figure \ref{fig:greenbox}.  Also depicted by the green rectangle is the global identification interval, i.e., the intersection of the individual intervals.  The global identification interval is indeed narrow, ranging from 0.8 to 2.0.   Evaluating (\ref{eq-idef1}) for $\overline{\tau}>0$ can be done via quadratic programming; see Appendix for details.  Figure \ref{fig:greenbox} (lower panel) shows that, when $\overline{\tau}=0.5$, the global identification interval is much wider: 0.6 to 2.9.  
 
 In summary, depending on the the upper limit in $\tau$ and the range in ${\gamma}^{[k]}$ values, i.e., the ``heterogeneity of bias,''   it appears that data can indeed contribute substantial information about $\theta$.

\subsubsection{Example: the NELS88* dataset}
\label{sec:nels88star}
Returning now to our example with the NELS88 dataset, we illustrate the impact of measurement error by intentionally adding non-differential measurement error to the data as described in equation (\ref{eq:nondiff}) so as to corrupt the reading test scores for 8 out of the $K=13$ schools.  As such, the contaminated dataset, NELS88$^{*}$, has $k^{'}=5$ schools for which the data are ``clean.''  We set $\phi^{[k]}=0$, for $k=1,\ldots,5$;  and, for  $k=6,\ldots,13$, increasing values from 1 to 12: $\phi^{[6]}=1.00$, $\phi^{[7]}=2.57 , \ldots, \phi^{[12]}=10.43$  and $\phi^{[13]}=12.00$.  Table \ref{tab:nels88starA} lists the aggregate data obtained after adding measurement error to the reading scores.  We also list the values of $\phi^{[k]}$ and $\gamma^{[k]}$ for reference.

We will fit both the univariate and bivariate \textit{BMEMA} models.  For the univariate model, we suppose that data for $\lambda^{[k]*}$s are unavailable and place a uniform prior on the $\gamma^{[k]}$ parameters as in (\ref{eq:uniform}).  We also fit an additional bivariate \textit{BMEMA} model with $\delta=0.5$ instead of $\delta=0.1$ (for all analyses where $K \ne k^{'}$) to see how sensitive results may be to the chosen priors.

\begin{table}[p]
\centering  
  \begin{tabular}{rccc|cc|rrrrr}
\multicolumn{8}{l}{\textbf{The NELS88$^{*}$ aggregate data} } &\\   
 &sample & measurement &  attenuation & \multicolumn{2}{l|}{data required for}  &  \multicolumn{4}{l}{data required for}   \\ 
 &size&error& factor  & \multicolumn{2}{l|}{univariate model}  &  \multicolumn{4}{l}{bivariate model}   \\ \hline
   $k$ & $n^{[k]}$  & $\phi^{[k]}$ & $\gamma^{[k]}$& $\hat{\beta}^{[k]*}$ & $\textrm{se}(\hat{\beta}^{[k]*})$ & $\hat{\alpha}^{[k]*}$ & $\textrm{se}(\hat{\alpha}^{[k]*})$  & $\hat{\beta}^{[k]*}$ & $\textrm{se}(\hat{\beta}^{[k]*})$ & $\hat{\sigma}^{[k]2*}$\\
  \hline
1 & 45 & 0.00 & 1.00 & 0.64 & 0.10 & 4.99 & 1.53 & 0.64 & 0.10 & 12.28 \\ 
  2 & 64 & 0.00 & 1.00 & 0.42 & 0.09 & 7.91 & 1.16 & 0.42 & 0.09 & 19.50 \\ 
  3 & 47 & 0.00 & 1.00 & 0.51 & 0.16 & 8.46 & 2.71 & 0.51 & 0.16 & 11.48 \\ 
  4 & 45 & 0.00 & 1.00 & 0.49 & 0.09 & 2.95 & 1.19 & 0.49 & 0.09 & 15.71 \\ 
  5 & 45 & 0.00 & 1.00 & 0.67 & 0.10 & 1.75 & 1.07 & 0.67 & 0.10 & 15.15 \\ 
  6 & 59 & 1.00 & 0.98 & 0.55 & 0.09 & 5.39 & 1.25 & 0.55 & 0.09 & 19.32 \\ 
  7 & 56 & 2.57 & 0.78 & 0.41 & 0.10 & 7.75 & 1.56 & 0.41 & 0.10 & 19.03 \\ 
  8 & 45 & 4.14 & 0.41 & 0.18 & 0.12 & 13.12 & 2.14 & 0.18 & 0.12 & 14.56 \\ 
  9 & 51 & 5.71 & 0.23 & 0.16 & 0.06 & 14.63 & 1.23 & 0.16 & 0.06 & 9.45 \\ 
  10 & 67 & 7.29 & 0.26 & 0.14 & 0.07 & 15.16 & 1.24 & 0.14 & 0.07 & 20.80 \\ 
  11 & 48 & 8.86 & 0.25 & 0.15 & 0.06 & 8.96 & 0.95 & 0.15 & 0.06 & 18.12 \\ 
  12 & 45 & 10.43 & 0.21 & 0.12 & 0.08 & 13.24 & 1.29 & 0.12 & 0.08 & 30.29 \\ 
  13 & 47 & 12.00 & 0.12 & 0.00 & 0.04 & 10.87 & 0.89 & 0.00 & 0.04 & 20.35 \\ 
   \hline
  \end{tabular}
\caption{The NELS88$^{*}$ data- Aggregate data obtained from the NELS88$^{*}$ dataset required for the \textit{BMEMA} univariate and bivariate models in \ref{sec:nels88star}. Values of  $\phi^{[k]}$ which correspond to the amount of measurement error intentionally added to each study and values of the attenuation factor, $\gamma^{[k]}$, are also listed for reference.}
\label{tab:nels88starA}
\end{table}


\begin{table}[ht]
\centering
\begin{tabular}{rcrrrlll}
  \hline
 &   {dataset}& model & $K$ & $k^{'}$  & $\theta$ & $CI(\theta)_{95\%}$   & $\tau$ \\
  \hline
 line 1. & NELS88$^{*}$ & univariate & 13&13 & 0.33 & 0.19 , 0.48 & 0.23 \\ 
 & & bivariate   &13&13 & 0.34 & 0.21 , 0.48 & 0.21 \\ 
line 2.& NELS88  & univariate &13&13 & 0.56 & 0.49 , 0.62 & 0.04 \\ 
 & & bivariate &13&13& 0.57 & 0.51 , 0.64 & 0.04 \\
 line 3. &NELS88$^{*}$  & univariate & 13&5 & 0.52 & 0.38 , 0.65 & 0.10 \\ 
 & & bivariate ($\delta = 0.1$)& 13&5 & 0.56 & 0.43 , 0.69 & 0.07 \\ 
 & & bivariate ($\delta = 0.5$)& 13&5 & 0.54 & 0.40 , 0.65 & 0.07 \\ 
 line 4.& NELS88$^{1:5}$  & univariate&5& 5 & 0.54 & 0.38 , 0.72 & 0.09 \\ 
 & & bivariate  & 5& 5 & 0.55 & 0.41 , 0.72 & 0.08 \\
 line 5. & NELS88$^{*}$ & univariate & 13&0  & 0.69 & 0.46 , 1.04 & 0.16 \\ 
 & & bivariate  ($\delta = 0.1$) & 13&0& 0.94 & 0.67 , 1.22 & 0.08 \\
 & & bivariate ($\delta = 0.5$) & 13&0& 0.74 & 0.48 , 1.04 & 0.08 \\ 
 line 6. & NELS88 & univariate & 13&5 & 0.63 & 0.54 , 0.77 & 0.09 \\ 
 & & bivariate ($\delta = 0.1$) & 13&5& 0.58 & 0.51 , 0.66 & 0.04 \\  
 & & bivariate ($\delta = 0.5$) & 13&5& 0.58 & 0.51 , 0.66 & 0.04 \\ 
 line 7. & NELS88 & univariate& 13&0 & 0.81 & 0.65 , 1.17 & 0.08 \\ 
 & & bivariate ($\delta = 0.1$) & 13&0& 0.60 & 0.52 , 0.78 & 0.05 \\ 
  & & bivariate ($\delta = 0.5$) & 13&0& 0.60 & 0.52 , 0.79 & 0.05 \\
   \hline
\end{tabular}
\caption{The data analysis results obtained - posterior medians with 95\% equal-tailed credible intervals. R-code to replicate table: \url{https://tinyurl.com/2ayeyeem}.}
\label{tab:nels88starB}
\end{table}

With the NELS88$^{*}$ dataset, we have that $\textrm{E}(\gamma^{[k]}) = 0.63$ and  $\operatorname{Var}(\gamma^{[k]}) = 0.15$.  Based on equation (\ref{eq:theta_me}), we have that ${\theta^{*}} =  0.66 \times 0.57 = 0.36$; and based on equation (\ref{eq:tau_me}), we have that ${\tau^{*2}} = {0.54}^{2}\times 0.04^{2}+ 0.15\times \left(0.04^{2}+ 0.57^{2}\right) = 0.05$, or $\tau^{*}=0.22$.  Indeed, if we ignore the possibility of any measurement error, we obtain, with the NELS88$^{*}$ dataset, estimates similar to the numbers given by equations (\ref{eq:theta_me}) and (\ref{eq:tau_me}) (see Table \ref{tab:nels88starB}  line 1):  when measurement error is added to the data, estimates for $\theta$ are biased downwards, while estimates for $\tau$ are biased upwards.  


  \begin{figure}
      \centering
      \includegraphics[width=17cm]{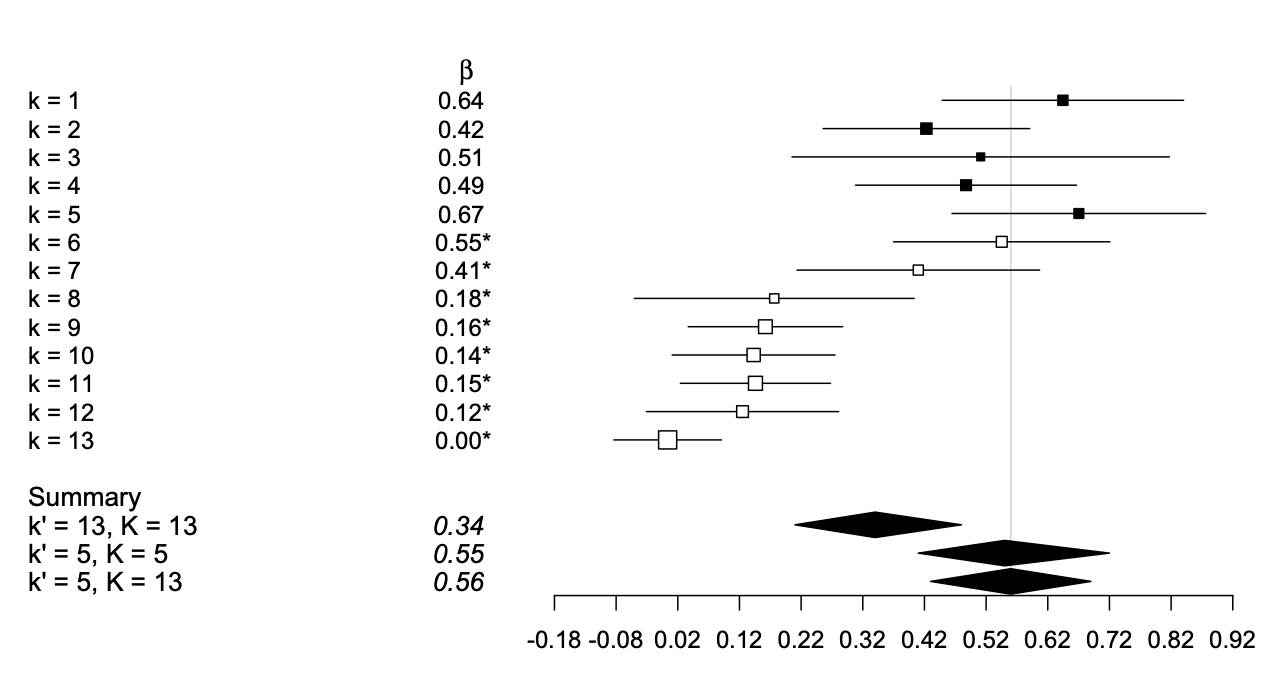}
      \caption{Forest plot for the meta-analysis of the NELS88$^{*}$ aggregate data, with three summaries corresponding to line 1 (in which all studies are assumed to be measurement error-free; $k^{'}=13$, $K=13$), line 4 (in which only the 5 studies known to be measurement error-free are included in the analysis; $k^{'}=5$, $K=5$), and line 3 (in which the 5 studies known to be measurement error-free are included in the analysis along with the 8 studies comprised by measurement error; $k^{'}=5$, $K=13$)  of Table \ref{tab:nels88starB}. The 5 black full squares correspond to the $\hat{\beta}^{[k]}$ values unaffected by measurement error; the 8 empty squares correspond  to the $\hat{\beta}^{[k]}$ values compromised by measurement error;  with horizontal lines corresponding to symmetrical $1.96\times\textrm{se}(\hat{\beta}^{[k]})$ confidence intervals.  The \textit{ BMEMA} posterior estimates of $\theta$ are plotted as diamonds, the lateral points of which indicate the equal-tailed 95\% credible intervals (i.e., the 2.5\% and 97.5\% quantiles) for these estimates.}
      \label{fig:forest_plot}
  \end{figure}

 For the NELS88$^{*}$ data, $\textrm{cor}(\hat{\beta}^{*[k]}, \hat{\lambda}^{*[k]}) = -0.66$, whereas  for the NELS88 data, \linebreak $\textrm{cor}(\hat{\beta}^{[k]}, \hat{\lambda}^{[k]}) = -0.02$.  The fact that the values of $\hat{\beta}^{*[k]}$ and $\hat{\lambda}^{*[k]}$ are negatively correlated in the presence of heterogeneous measurement error and independent otherwise suggests that by combining data from several studies, the presence of bias caused by measurement error can be better identified (and suggests the possibility of a simple diagnostic test for heterogeneous measurement error).

\begin{figure}
    \centering
    \includegraphics[width=12cm]{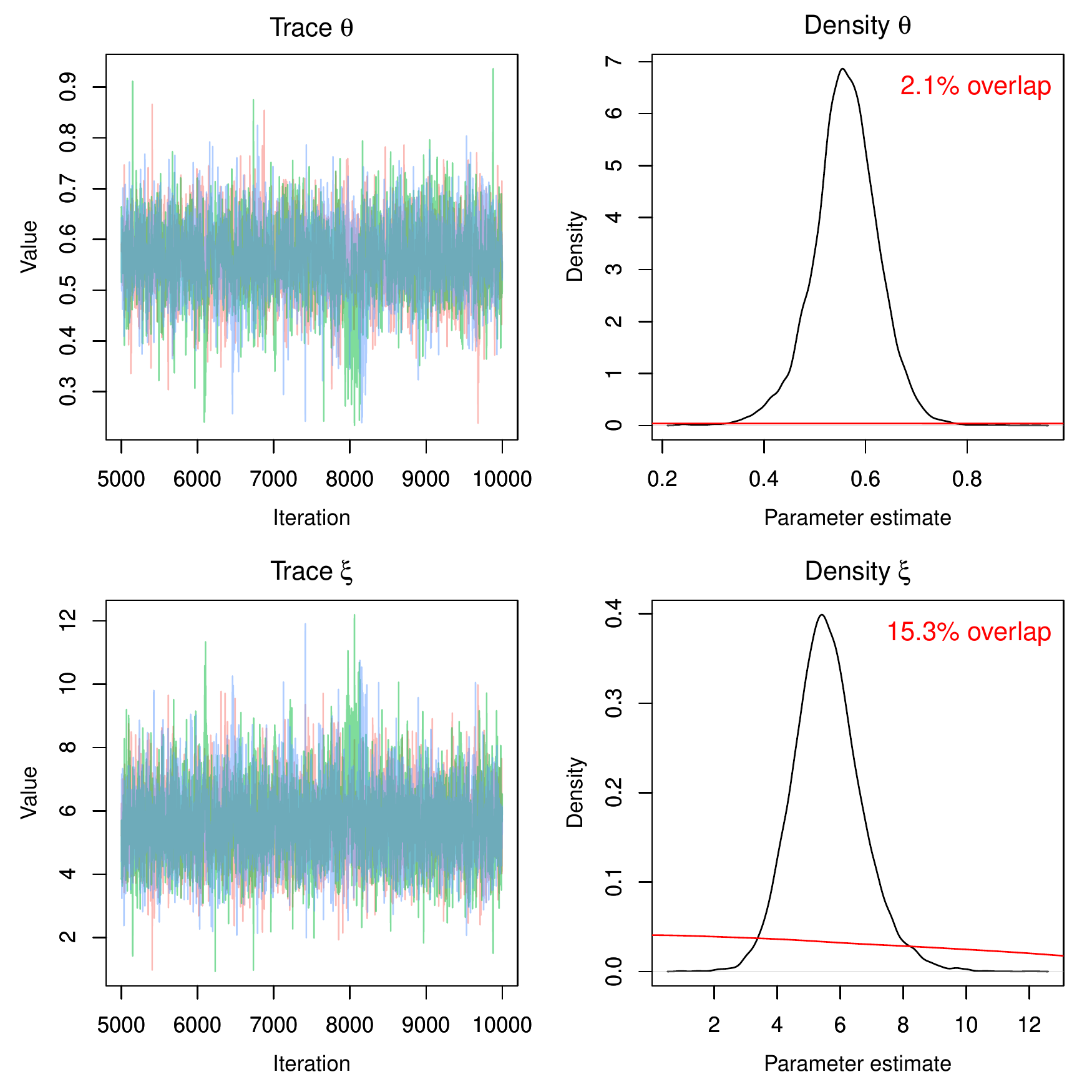}
    \caption{Diagnostic plots for parameters $\theta$ and $\xi$, for the MCMC simulation of the univariate \textit{ BMEMA} model with $k^{'} = 5$ and $\delta=0.1$ (this corresponds to the results in Table \ref{tab:nels88starB}, line 3). The left panels report trace plots from the posterior to check convergence. The right panels report the corresponding posterior distribution estimate (black solid line) together with the prior distribution for that parameter (red solid line). The \% overlap reported in red is the PPO (prior-posterior overlap).}
    \label{fig:MCMC1}
\end{figure}

\begin{figure}
    \centering
    \includegraphics[width=12cm]{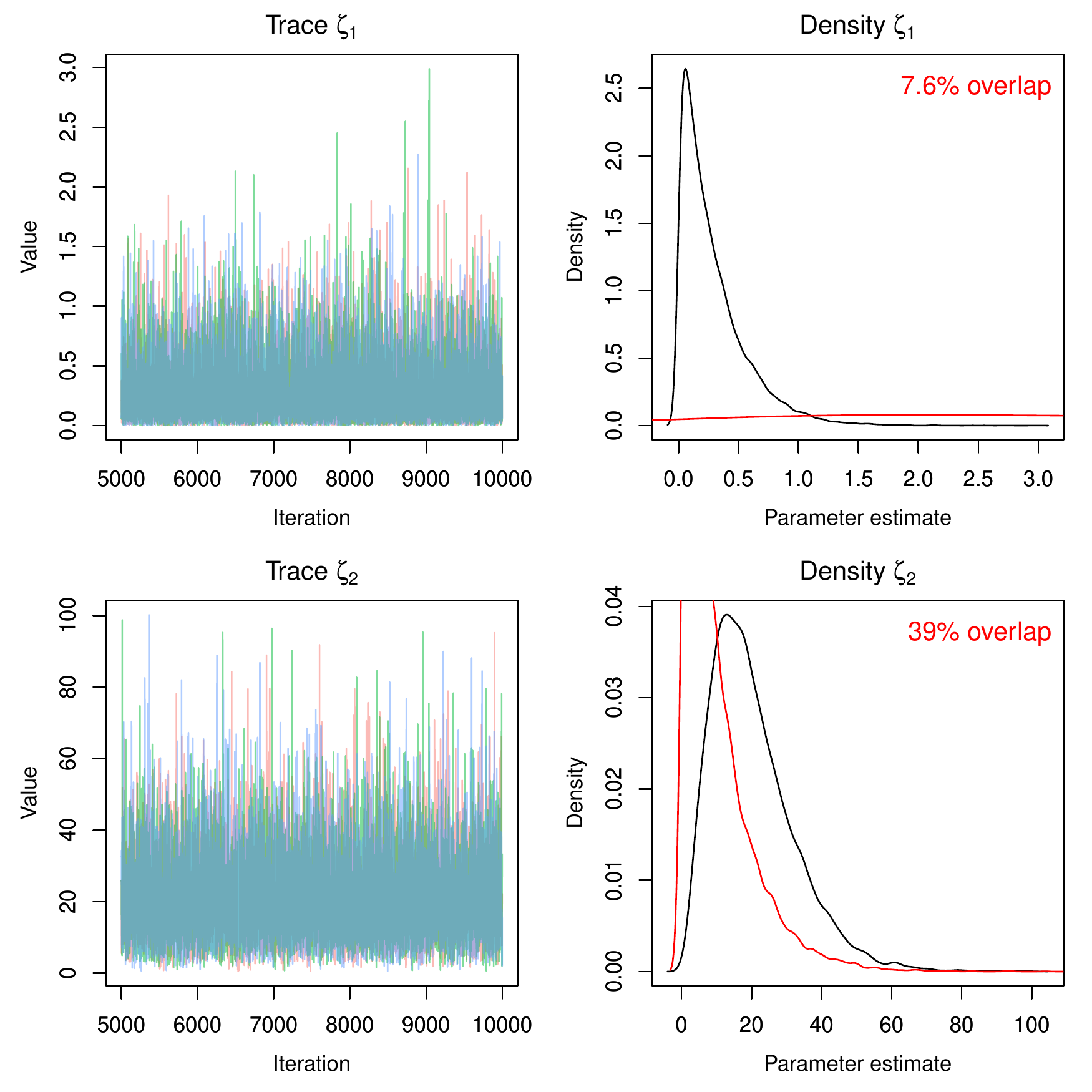}
    \caption{Diagnostic plots for parameters $\zeta_{1}$ and $\zeta_{2}$, for the MCMC simulation of the univariate \textit{ BMEMA} model with $k^{'} = 5$ and $\delta=0.1$ (this corresponds to the results in Table \ref{tab:nels88starB}, line 3). The left panels report trace plots from the posterior to check convergence. The right panels report the corresponding posterior distribution estimate (black solid line) together with the prior distribution for that parameter (red solid line). The \% overlap reported in red is the PPO (prior-posterior overlap).}
    \label{fig:MCMC2}
\end{figure}

\textcolor{black}{The estimates of $\hat{\theta}=0.56$ and $\hat{\theta}=0.57$ obtained with the univariate and bivariate \textit{ BayesMA} models respectively, with the ``clean'' NELS88 dataset (see Table \ref{tab:nels88starB}, line 2) serve as approximate targets.  For the NELS88$^{*}$ dataset, the bivariate \textit{BMEMA} model, with $\delta=0.1$, and with $k^{'}=5$, obtains an estimate of $\hat{\theta} = 0.56$, with 95\% credible interval of $CI(\theta)_{95\%} = [0.44, 0.68]$ (see Table \ref{tab:nels88starB}, line 3).  MCMC diagnostic plots are presented in the Appendix and show little prior-posterior overlap (PPO) which suggests that the prior is suitably overwhelmed by the signal provided in the data; see Figures \ref{fig:MCMC1} and \ref{fig:MCMC2}.  With the alternative prior specified by $\delta=0.5$, the bivariate model obtains an estimate of $\hat{\theta} = 0.54$.  The univariate model obtains an estimate of $\hat{\theta} = 0.52$, with 95\% credible interval of $CI(\theta)_{95\%} = [0.39, 0.66]$.}

\textcolor{black}{We also fit the Bayesian model to data from only the first five schools that are known to be unaffected by measurement error  (see Table \ref{tab:nels88starB}, line 4) and obtain, with the bivariate model, $\hat{\theta}=0.55$ with a notably wider credible interval of $CI(\theta)_{95\%} = [0.41, 0.73]$ (and similar results with the univariate model).  This suggests that the data from the additional eight schools, while compromised by measurement error, may provide ``added value'' and sharpen our inference. }

\textcolor{black}{We also note that knowing that the $k=1,\ldots,5$ schools are untainted by measurement error is crucial to obtaining appropriate estimates: the \textit{ BMEMA} models, with $k^{'}=0$, obtain estimates of $\theta$ much too high: $\hat{\theta}=0.68$ (for the univariate), 0.94 (for the bivariate with $\delta=0.1$), and 0.75 (for the bivariate with $\delta=0.5$) (see Table \ref{tab:nels88starB}, line 5).  These three estimates are quite different and this suggests that, when there are no known ``gold standard studies'' to ``anchor'' the estimates, the chosen priors may yield significant leverage.}

Finally, note that if the \textit{ BMEMA} model is fit to the original NELS88 data, we will end up slightly overestimating the $\theta$ parameter.  For the bivariate model (with $\delta=0.1$), we obtain $\hat{\theta}=0.58$, with $k^{'}=5$, and $\hat{\theta}=0.61$, with $k^{'}=0$ (see Table \ref{tab:nels88starB}, lines 6 and 7).

\section{Meta-analysis of multivariable linear regression}

\subsection{In the absence of measurement error}
\label{sec:multiwme}

 \textcolor{black}{We can generalize the \textit{ BayesMA} models described in Section \ref{sec:linmodel} (in which each study can be summarized as a simple linear regression) to a general case where each study can be summarized as a multivariable linear regression.  While this generalization is of theoretical interest, in practice it may be unlikely to have multiple different studies provide results from exactly the same regression model.  Should different studies adjust for different subsets of covariates, pooling their coefficients together in a meta-analysis may not be appropriate \cite{greenland1987quantitative}.  However, note that meta-analytic methods to synthesize linear regressions with different covariates have been developed; see for example Yoneoka et al. (2017) \cite{yoneoka2017synthesis} and Debray et al. (2012) \cite{debray2012incorporating}.}

 We denote \(X_{j \times}^{[k]}\) as the $(Q+1)$-length row vector with elements 1, and the $Q$ covariates measured for the \(j\)-th observation in the \(k\)-th observational study.  Similarly, we denote \(X_{\times q}^{[k]}\) as the $n^{[k]}$-length column vector of values of the $q$-th covariate for the $k$-th study.  Finally the $n^{[k]} \times (Q+1)$ design matrix, \(X^{[k]}\), consists of $n^{[k]}$ rows, one for each observation in the $k$-th study: \(X_{j \times}^{[k]}\), for $j$ in $1,\ldots,n^{[k]}$.  Note that, in this multivariate setting, \(X^{[k]}\) is a matrix with the first column consisting of 1s.   We will use $X_{\times, -1}^{[k]}$ to denote the design matrix excluding the column of 1s.

Suppose data from each study can be summarized as a multivariable linear regression model such that, for $k = 1,\ldots,K$, and $j$ in $1,\ldots,n^{[k]}$:
\begin{equation}
Y_{j}^{[k]} | X_{j \times}^{[k]} \sim  \mathcal{N}\left(X_{j \times}^{[k]}\beta^{[k]}, \sigma^{[k]2}\right),
\label{eq:Yj}
\end{equation}
\noindent where $\beta^{[k]} = (\beta_{0}^{[k]}, \beta_{1}^{[k]},\ldots, \beta_Q^{[k]})^{'}$ is the column-vector of regression coefficients.  We assume that each study has its own exposure distribution governed by a multivariate Normal distribution:
\begin{equation}
X_{j, -1}^{[k]} \sim \mathcal{MVN}\left(\mu^{[k]}, \Lambda^{[k]}\right),
\label{eq:Xj}
\end{equation}
\noindent where $\mu^{[k]}$ is a $Q$-length vector, and $\Lambda^{[k]}$ is a $Q \times Q$ covariance matrix. 
The random-effects meta-analysis model can be summarized as:
\begin{equation}
\beta^{[k]}| \theta, \Tau \sim  \mathcal{MVN}(\theta, \Tau) , \label{eq:mvn2}
\end{equation} 
\noindent for all $k = 1,\ldots,K$; where $\theta$ is a $(Q+1)$-length  vector, and $\Tau$ is a $(Q+1) \times (Q+1)$ covariance matrix.  If we are able to assume that the regression coefficients are \textit{a priori} independent, then $\Tau$ will be a diagonal matrix. 

If individual participant data (IPD) are available, the model will have many moving parts.  For our unknown parameters of interest ($\theta$, $\Tau$, $\beta$, $\sigma$, $\mu$, $\Lambda$, and $\Tau$), and the data from $K$ studies (we require: $Y_{j}^{[k]}$, and $X_{j \times}^{[k]}$ for $j$ in $1,\ldots,n^{[k]}$, and for $k = 1,\ldots,K$), Bayes' theorem states that:
\begin{align}
    p( \theta, \Tau, \beta, \sigma, \mu, \Lambda | \textrm{data}) &\propto p(\textrm{data}|( \theta, \Tau, \beta, \sigma, \mu, \Lambda))p( \theta, \Tau, \beta, \sigma, \mu, \Lambda) \nonumber \\
    = \prod_{k=1}^{K}  \Big( \prod_{j=1}^{n^{[k]}}&\Big\{p(Y_{j}^{[k]}| X_{j \times}^{[k]}, \beta^{[k]}, \sigma^{[k]}) p(X_{j \times}^{[k]} | \mu^{[k]}, \Lambda^{[k]}) \Big\} \nonumber \\
    & \quad \quad \times p(\beta^{[k]}|\theta, \Tau) p(\mu^{[k]}) p(\Lambda^{[k]}) p(\sigma^{[k]}) \Big) p(\theta) p(\Tau).
\label{eq:bayesMulti}
\end{align}


If IPD are not available, and only aggregate data are available, a simpler two-part model can be defined whereby: 
 \begin{equation}
 \hat{\beta}^{[k]}|\beta^{[k]} \sim \mathcal{MVN}(\beta^{[k]} , COV^{[k]}) \textrm{, where  } COV^{[k]} = (X^{[k]T}X^{[k]})^{-1}\sigma^{[k]2}; \textrm{and}
 \label{eq:mvn1}
 \end{equation}
 \begin{equation}
\beta^{[k]}| \theta, \Tau \sim  \mathcal{MVN}(\theta, \Tau) , \label{eq:mvn2}
\end{equation} 

 \noindent  for $k = 1,\ldots,K$; where $COV^{[k]}$ is assumed available and known for $k=1,\ldots,K$ (this is analogous to the assumption in Section \ref{sec:linmodel} that,  for $k = 1,\ldots,K$, $\mu^{[k]}$, $\sigma^{[k]}$, and $\lambda^{[k]}$ are known).
%
%

\subsection{Adjusting for non-differential measurement error}

Suppose now that observed covariates are measured with non-differential error such that:
\begin{equation}
X_{j, -1}^{*[k]} =  \mathcal{MVN}(X_{j, -1}^{[k]} , \Phi^{[k]}),
\label{eq:multiME}
\end{equation}
\noindent for  $j$ in $1,\dots,n^{[k]}$ and for $k = 1,\ldots,K$; where $\Phi^{[k]}$ is a $Q\times Q$ covariance matrix.  Note that, equation (\ref{eq:multiME}) is analogous to equation (\ref{eq:nondiff}) in Section \ref{sec:univariate_withME}.  Also note that, if $\Phi^{[k]}$ is a diagonal matrix, then the measurement error in any one covariate is entirely independent of the measurement error in every other covariate.  

Multivariate measurement error can bias estimators in unpredictable and unexpected ways.  For instance, Abel (2017) \cite{abel2017classical} shows that, even if the $q$-th covariate, $X_{\times q}$, is measured without error, one may still incorrectly reject the null hypothesis of $\beta_{q}^{[k]}=0$, if $X_{\times q}$ is correlated with another covariate that is itself tainted by measurement error.  Since covariates may be correlated to one another in many different ways, it is difficult to anticipate the impact of multivariate measurement error for general multivariate settings \cite{levi1973errors}.

The multivariate \textit{ BayesMA} model outlined in Section \ref{sec:multiwme} can be adapted to account for measurement error in one or several of the covariates.  If IPD are available \cite{debray2015get}, we can frame a flexible multivariate \textit{ BMEMA} model by assuming that the covariates are multivariate normal (but this could be modified as needed) and defining the following three-part model structure:
\begin{align}
    X_{j, -1}^{[k]} &\sim \mathcal{MVN}\left(\mu^{[k]}, \Lambda^{[k]}\right), \label{eq:threeX}\\
    X_{j, -1}^{[k]*} | X_{j, -1}^{[k]} &\sim \mathcal{MVN}\left(X_{j, -1}^{[k]} , \Phi^{[k]}\right),  \label{eq:threeXstar}\\
 \quad \textrm{and} \quad   Y_{j}^{[k]} | X_{j \times }^{[k]} &\sim  \mathcal{N}\left(X_{j \times }^{[k]}\beta^{[k]}, \sigma^{[k]2}\right), \label{eq:threeY}
\end{align}
 \noindent for $j$ in $1,\ldots,n^{[k]}$ and $k = 1,\ldots,K$.

%

  As a prior for $\Phi$,  the inverse-Wishart distribution with covariance matrix $2\zeta_{2}\times I_{[Q]}$ and $2\zeta_{1}$ degrees of freedom  is a multivariate generalization of the prior we considered in Section \ref{sec:univariate_withME} for $\phi^{k]}$.  Consider:
 \begin{equation}
     \Phi^{[k]} \sim \textrm{Inv-Wishart}(2\zeta_{2}\times I_{[Q]}, 2\zeta_{1});  \quad \quad \zeta_{1} \sim \textrm{Exp}(\delta) \quad \textrm{and:} \quad \zeta_{2} \sim \textrm{Exp}(\delta),
 \end{equation}
 \noindent for $k = 1,\ldots,K$; where $\zeta_{1}>(Q/2)$ and $\zeta_{2}>0$.  Indeed, if $Q=1$, we have that, \textit{ a priori},  $\textrm{E}(\Phi^{[k]}) = 2\zeta_{2}/(2\zeta_{1} - Q - 1) = \zeta_{2}/(\zeta_{1} -1)$; and $\textrm{Var}(\Phi^{[k]}) = \zeta_{2}^{2}/((\zeta_{1} - 1)^{2}(\zeta_{1} - 2))$.

\subsection{Example: the NELS88 dataset}

We return to the NELS88 dataset example and consider $Q=2$ covariates.  Let $X_{1}$ be the reading test score and $X_{2}$ be the mathematics test score.  In order to illustrate the impact of measurement error, we will create the NELS88$^{*}$ dataset by adding non-differential measurement error as described in equation  (\ref{eq:multiME}) so as to corrupt both the reading test scores and the mathematics test score for 8 out of the $K=13$ schools.  As such, as in Section \ref{sec:nels88star}, we have $k^{'}=5$.  We define $\Phi^{[k]}$ to be a diagonal matrix such that the measurement error in $X_{1}$ is independent of the measurement error in $X_{2}$. For $k=6,\ldots,13$, we set $\Phi_{1,1}^{[k]}$ equal to between 4 and 6; and set $\Phi_{2,2}^{[k]}$ equal to between 8 and 12.  Table \ref{tab:nels88starmulti} - A lists regression coefficients obtained before and after adding the measurement error and also lists the set values for $\Phi^{[k]}_{1,1}$ and $\Phi^{[k]}_{2,2}$ for reference.

For this simple example analysis, we specify multivariate normal priors for $\mu^{[k]}$ and $\theta$: $\mu^{[k]} \sim \mathcal{MVN}(0_{[Q]}, 100 \times I_{[Q]})$,  for $k = 1,\ldots,K$; and $\theta \sim \mathcal{MVN}(0_{[Q+1]}, 100 \times I_{[Q+1]});$ \noindent where $0_{[Q]}$ is a $Q$-length vector of zeros, and $I_{[Q]}$ is a $Q \times Q$ identity matrix.  Also, we will assume that the regression coefficients are \textit{a priori} independent and specify half-Cauchy priors for $ \sigma^{[k]}$ and the $Q+1$ diagonal elements of $\Tau$: $\sigma^{[k]} \sim \textrm{half-Cauchy}(0, 2), $ for $k = 1,\ldots,K$; and $\sqrt{\Tau_{qq}} \sim \textrm{half-Cauchy}(0, 2), $ for $q$ in $1,\ldots,Q+1$.  Finally, for $\Lambda^{[k]}$ and $\Phi$, we specify inverse-Wishart priors such that: $\Lambda^{[k]} \sim \textrm{Inv-Wishart}(I_{[Q]}, Q), $ for $k = 1,\ldots,K$; and $ \Phi^{[k]} \sim \textrm{Inv-Wishart}(2\zeta_{2}\times I_{[Q]}, 2\zeta_{1});   \quad \zeta_{1} \sim \textrm{Exp}(0.1), \quad \textrm{and} \quad \zeta_{2} \sim \textrm{Exp}(0.1)$.

Table \ref{tab:nels88starmulti} - B lists parameter estimates obtained from the multivariate \textit{ BMEMA} model.  The measurement error introduced to the data biases the estimate of $\theta_{2}$ towards 0.  With the unbiased data, and $k^{'}=13$, we obtain $\hat{\theta}_{2} = 0.34$ (see Table \ref{tab:nels88starmulti}-B, line 2).  In contrast, with the biased data, NELS88$^{*}$, we obtain $\hat{\theta}_{2} = 0.25$  (see Table \ref{tab:nels88starmulti}-B, line 1).  The \textit{ BMEMA} model obtains a an estimate of $\hat{\theta}_{2} = 0.35$ (see Table \ref{tab:nels88starmulti}-B, line 3, and MCMC diagnostic plots in Figure \ref{fig:MCMCmulti_line3} in the Appendix).  When $k^{'}=0$, the MCMC mixing is problematic; this is clear in the MCMC diagnostic plots; see Figure \ref{fig:MCMCmulti_line5} in the Appendix.  The challenging sampling is no doubt due to to the identifiability issues discussed in Section \ref{sec:unident} and to the fact that different combinations of $\Phi^{[k]}$, $\theta^{[k]}$ and $\Tau^{[k]}$ can yield similar model probabilities. Unless custom samplers are configured, inference from the \textit{ BMEMA} model with $k^{'}=0$ may not be possible.  This should not be so surprising: computation with ``partially identified'' models can be a ``bottleneck issue'' (see Section 7.1 of Gustafson (2015) \cite{gustafson2015bayesian}).

\begin{table}[p]
\centering
\begin{tabular}{lccclclclcc}
\hline
\multicolumn{7}{l}{\textbf{A. NELS88$^{*}$ data} }  & & & \\   
  $k$ & & $n^{[k]}$ & $\hat{\beta}_{1}^{[k]}$ & $\hat{\beta}_{1}^{[k]*}$ & $\hat{\beta}_{2}^{[k]}$ &   $\hat{\beta}_{2}^{[k]*}$ & $\hat{\beta}_{3}^{[k]}$ & $\hat{\beta}_{3}^{[k]*}$   & $\sqrt{\Phi_{1,1}^{[k]}}$ & $\sqrt{\Phi_{2,2}^{[k]}}$\\ \hline
1 && 45 & 4.38 & 4.38 & 0.39 & 0.39 & 0.18 & 0.18 & 0.00 & 0.00 \\ 
  2 && 64 & 5.47 & 5.47 & 0.26 & 0.26 & 0.22 & 0.22 & 0.00 & 0.00 \\ 
  3 && 47 & 4.31 & 4.31 & 0.28 & 0.28 & 0.26 & 0.26 & 0.00 & 0.00 \\ 
  4 && 45 & 2.35 & 2.35 & 0.20 & 0.20 & 0.18 & 0.18 & 0.00 & 0.00 \\ 
  5 && 45 & 0.23 & 0.23 & 0.34 & 0.34 & 0.28 & 0.28 & 0.00 & 0.00 \\ 
  6 && 59 & 3.59 & 6.70 & 0.27 & 0.19 & 0.25 & 0.14 & 6.00 & 12.00 \\ 
  7 && 56 & 2.29 & 4.53 & 0.31 & 0.31 & 0.29 & 0.20 & 6.00 & 8.00 \\ 
  8 && 45 & 3.60 & 15.94 & 0.26 & 0.04 & 0.25 & -0.01 & 6.00 & 10.00 \\ 
  9 && 51 & 2.16 & 14.20 & 0.50 & 0.18 & 0.19 & 0.01 & 5.00 & 10.00 \\ 
  10 && 67 & 5.62 & 8.32 & 0.64 & 0.30 & 0.04 & 0.14 & 6.00 & 8.00 \\ 
  11 && 48 & 3.62 & 5.87 & 0.41 & 0.25 & 0.13 & 0.08 & 5.00 & 10.00 \\ 
  12 && 45 & 3.14 & 7.62 & 0.38 & 0.34 & 0.25 & 0.11 & 5.00 & 12.00 \\ 
  13 && 47 & 3.78 & 6.37 & 0.15 & 0.06 & 0.25 & 0.19 & 5.00 & 8.00 \\ 
  
     \hline
   \vspace{0.05cm} \\
  \hline   
\multicolumn{9}{l}{\textbf{B. Analysis results with \textit{ BMEMA}, $Q=2$} }&    \\   
\hline
 &  {dataset} & $K$, $k^{'}$ & $ \theta_{1}$ & $CI(\theta_{1})$ & $\theta_{2}$ & $CI(\theta_{2})$ &  $\theta_{3}$ & $CI(\theta_{3})$ &  $\sqrt{\Tau_{22}}$ & $\sqrt{\Tau_{33}} $\\
\multicolumn{2}{l}{line  1. NELS88$^{*}$ }& 13, 13& 6.17 & 4.49, 7.99 & 0.25 & 0.19, 0.32 & 0.15 & 0.11, 0.20 & 0.05 & 0.06 \\ 
\multicolumn{2}{l}{line   2. NELS88 } & 13, 13 & 3.14 & 2.11, 4.26 & 0.34 & 0.26, 0.42 & 0.22 & 0.18, 0.27 & 0.07 & 0.02 \\ 
\multicolumn{2}{l}{ line  3. NELS88$^{*}$ }& 13, 5 & 2.41 & 0.86, 4.04 & 0.35 & 0.23, 0.48 & 0.25 & 0.18, 0.32 & 0.07 & 0.04 \\ 
 \multicolumn{2}{l}{line  4. NELS88$^{1:5}$ }& 5, 5 & 3.01 & 1.03, 5.35 & 0.30 & 0.13, 0.46 & 0.23 & 0.14, 0.33 & 0.08 & 0.05 \\ 
\multicolumn{2}{l}{ line  5. NELS88$^{*}$} & 13, 0 & -- & --, -- & -- & --, -- & -- & --, --  & -- & --\\ 
\multicolumn{2}{l}{line   6. NELS88 }& 13, 5 & 2.28 & 1.12, 3.47 & 0.36 & 0.25, 0.48 & 0.24 & 0.18, 0.31 & 0.07 & 0.03 \\ 
\multicolumn{2}{l}{line   7. NELS88 }& 13, 0 & -- & --, -- & -- & --, -- & -- & --, -- & -- & --\\ 
   \hline
\end{tabular}
\caption{ \textcolor{black}{A. Aggregate data obtained from the NELS88 dataset and the NELS88$^{*}$ dataset for comparison.  Values for $\Phi^{[k]}_{1,1}$ and $\Phi^{[k]}_{2,2}$ which correspond to the amount of measurement error intentionally added to the reading test score and the mathematics test score, respectively, for  the $k$-th study are also listed for reference.   B. The data analysis results obtained - posterior medians with 95\% equal-tailed credible intervals.  R-code to replicate the table: \url{https://tinyurl.com/y5w9jxg7}.}}
\label{tab:nels88starmulti}
\end{table}

\section{Conclusion}
 
 \textcolor{black}{
   A meta-analysis based on all available evidence, even if some evidence is less than perfect, may be preferable to a meta-analysis that ignores large swathes of data \cite{turner2009bias} \cite{stone2019comparison}.  However, one should always correct for sources of bias if this is possible.  Currently, tools to correct for measurement error in a meta-analysis are not available and as a consequence, researchers are left to simply list measurement error as a study limitation (e.g. Wu et al. (2016)\cite{wu2016meta}: ``this study has several limitations [...] none of the studies corrected for measurement error.''; and Merino et al. (2009) \cite{merino2019quality}: ``although every effort was made to maximize the validity of the study, minimize bias, and incorporate heterogeneity and uncertainty, the estimated hazard ratios of dietary components could be affected by measurement error'').}
  
  In the simplest univariate scenario, if the exposure of interest in a study is compromised due to non-differential measurement error, one must inflate its point estimates and down-weight the study's overall contribution to the meta-analysis.  The proposed Bayesian model, the \textit{ BMEMA} model, provides a systematic and efficient way to do just this for continuous outcome data.  If one suspects that certain studies are compromised by measurement error, one can incorporate this uncertainty regarding the bias into the hierarchical Bayesian framework and obtain appropriate point and interval estimates.  Moreover, as we demonstrated with the NELS88 analysis example, incorporating these biased studies can be beneficial: credible intervals were narrower when data from all studies were included for meta-analysis relative to when only the unbiased studies were included.  This is relevant in real-world settings where meta-analyses pool evidence from varied sources.  For example, in epidemiology, observational studies are frequently combined with randomized control trials (RCTs) in systematic reviews and meta-analyses; see Bun et al. (2020) \cite{bun2020meta}.  While certain studies may be biased, they may still provide value if one can appropriately account and adjust for the bias.  Bayesian inference is well suited to the task.

 \textcolor{black}{ We also showed that, if IPD are available, a Bayesian meta-analysis model can easily adjust for multiple participant-level covariates that are measured with or without measurement error \cite{higgins2001meta}.  While issues of identifiability may make a Bayesian model difficult to fit for more complex multivariable data, so long as a subset of studies is known to be unbiased, these studies can ``anchor'' the uncertainty allowing for straightforward Bayesian inference.  That being said, the proposed \textit{BMEMA} model requires one to make several important assumptions.  We stress that the \textit{BMEMA} model will assume that all differences between the unbiased studies and the biased studies are due to measurement error when in reality there may be other systematic differences at play.  Furthermore, the model, as it is currently described, can only deal with non-differential measurement.  In practice, there may be substantial bias due to differential measurement error\cite{van2020reflection}.  Differential measurement error is a major concern, for instance, in retrospective studies; see White (2003) \cite{white2003design}. Future research should investigate how to address these difficult issues.  In addition, future research should generalize the proposed \textit{BMEMA} model for binary and time-to-event outcomes and could also extend the model to network meta-analysis.}

\textcolor{black}{On a final note, beyond the bias caused by measurement error, a meta-analysis of observational studies should, ideally, also take into account other (potentially bigger) biases, e.g.:  publication bias and bias due to unmeasured confounding }\cite{maier2020robust,mccandless2007bayesian}). The solutions we put forward may be more broadly applicable and it would seem desirable, and feasible, to consider all sources of uncertainty and bias within a single comprehensive Bayesian model.  Future work should investigate whether the Bayesian hierarchical framework proposed and the ``heterogeneity of bias'' principle can be used to derive appropriate estimates in a meta-analysis where individual studies are subject to varying degrees --and varying types-- of bias.  

\vspace{1cm}

\section*{HIGHLIGHTS}

\paragraph{What is already known:}
\begin{itemize}
\item It is important to adjust for known sources of measurement error when conducting a meta-analysis.   If the exposure of interest in a study is compromised due to non-differential measurement error, one must simply inflate its point estimates and down-weight the study's overall contribution to the meta-analysis. 
\end{itemize}

\paragraph{What is new:}
\begin{itemize}
\item The proposed Bayesian model, the \textit{ BMEMA} model, provides a systematic and efficient way to conduct a meta-analysis of measurement error-tainted continuous outcome data.  If individual participant data (IPD) are available, a Bayesian meta-analysis model can adjust for multiple participant-level covariates that are measured with or without measurement error.
\end{itemize}

\paragraph{Potential impact for RSM readers outside the authors' field:}
\begin{itemize}
\item  Meta-analyses based on all available evidence, even if some evidence is less than perfect, may be preferable to meta-analyses that ignore large swathes of data.
\end{itemize}

\vspace{1cm}

\section*{DATA AVAILABILITY STATEMENT}  

Code to replicate all analysis in this paper is available in two R files at \url{https://github.com/harlanhappydog/MEMA}:

\begin{itemize}
\item univMEMAjags.R replicates all results in Table \ref{tab:nels88starB}.
\item multiMEMAjags.R replicates all results in Table \ref{tab:nels88starmulti}.
\end{itemize}

\bibliography{wileyNJD-AMA}  

\section{Appendix}

Consider the evaluation of (\ref{eq:partialid}) for $\overline{\tau}>0$, i.e., where a limited heterogeneity in the $\beta^{[k]}$ parameters is permitted.  Figure \ref{fig:greenbox} (lower panel) shows how the global identification interval is wider when $\overline{\tau}=0.50$ relative to when $\overline{\tau}=0$ for the ``fixed-effects'' case.  This interval outlined by the green rectangle can be easily obtained via quadratic programming. 

Recall that quadratic programming constitutes the minimization of a quadratic function subject to linear constraints, and these may be a mix of equality and inequality constraints.   Let $x$ be a candidate value, which we will test for membership in the identification interval.        
To perform this test, we use a standard quadratic programming package (e.g., quadprog \cite{weingesselquadprog})
to minimize the quadratic function $Var(\beta^{[k]})$, subject to the equality constraint $\theta=x$ and the inequality constraints which restrict $\beta^{[k]}$ to the interval $I_k$ for each $k$. The $x$ value belongs in the identification interval if and only if the minimized variance does not exceed $\overline{\tau}^{2}$.  Thus a simple grid search over values of $x$ numerically determines the identification interval.  Two numerical searches can be undertaken.  One starts at the underlying value and tests successively larger $x$ until a failing value is obtained.  The other starts at the underlying value and does the same, but moving downwards.

\begin{figure}
    \centering
    \includegraphics[width=12cm]{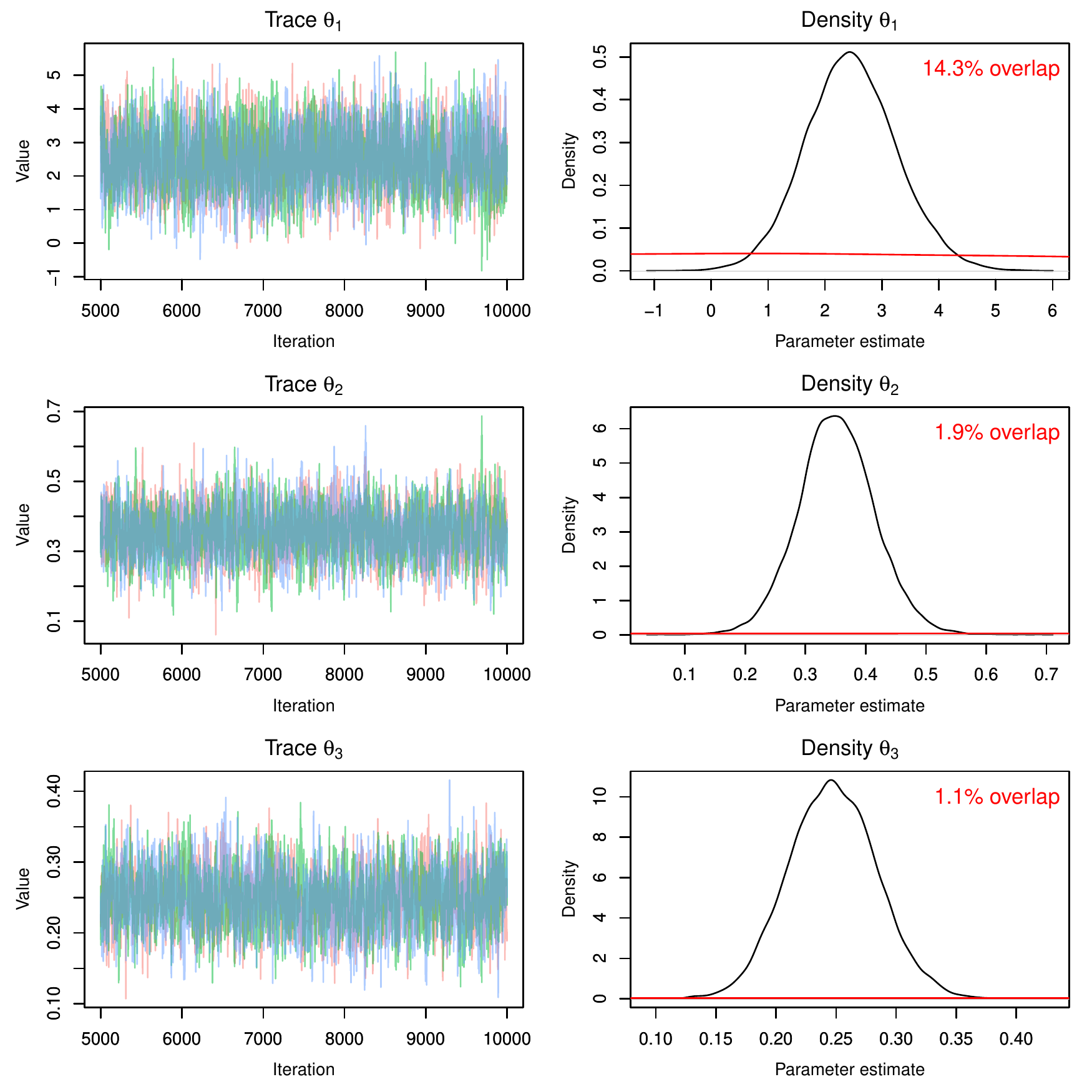}
    \caption{Diagnostic plots for the $\theta$ parameter, for the MCMC simulation of the multivariate \textit{ BMEMA} model with $k^{'} = 5$ (this corresponds to the results in Table \ref{tab:nels88starmulti}-B, line 3). The left panels report trace plots from the posterior to check convergence. The right panels report the corresponding posterior distribution estimate (black solid line) together with the prior distribution for that parameter (red solid line). The \% overlap reported in red is the PPO (prior-posterior overlap).}
    \label{fig:MCMCmulti_line3}
\end{figure}

\begin{figure}
    \centering
    \includegraphics[width=12cm]{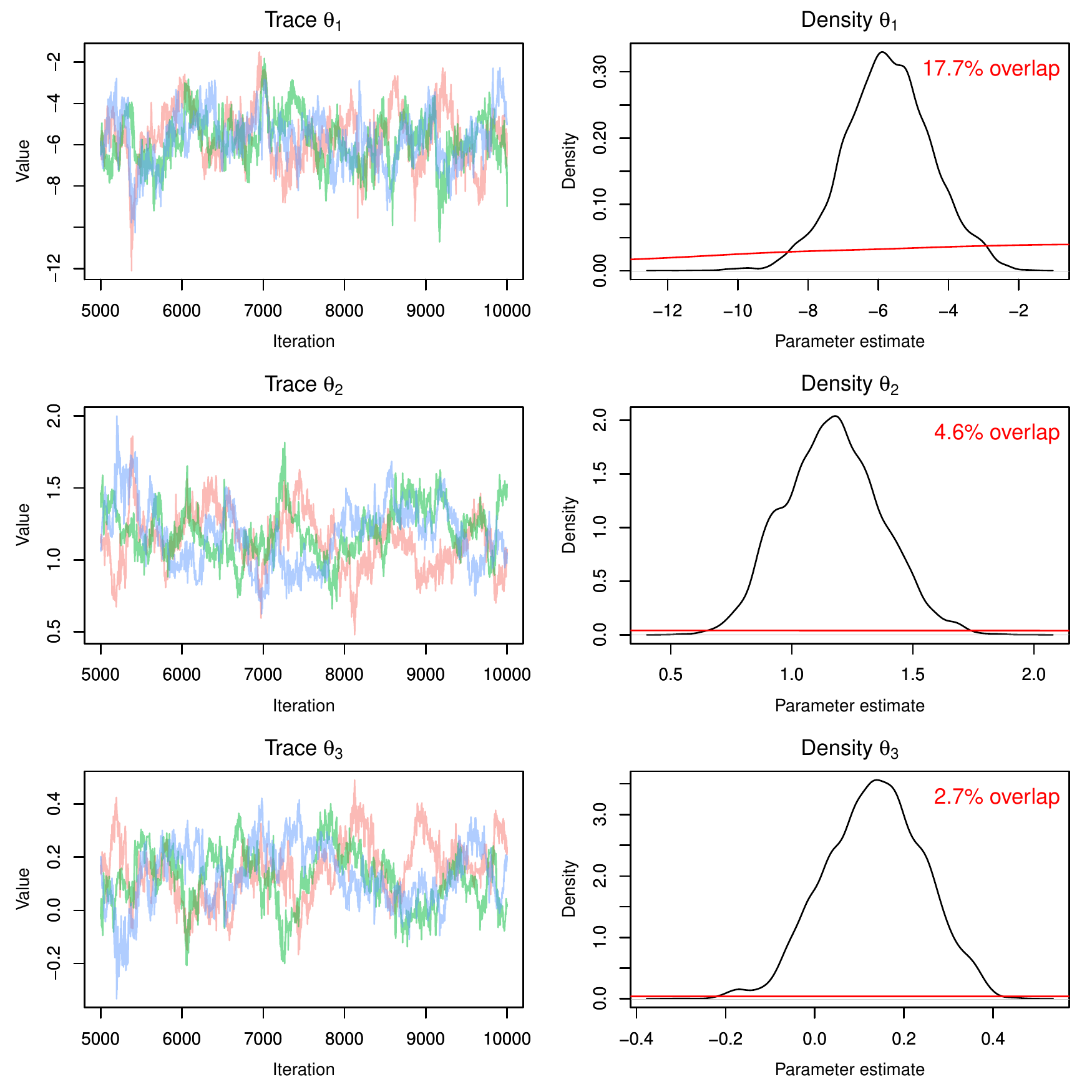}
    \caption{Diagnostic plots for the $\theta$ parameter, for the MCMC simulation of the multivariate \textit{ BMEMA} model with $k^{'} = 0$ (this corresponds to the results in Table \ref{tab:nels88starmulti}-B, line 5). The left panels report trace plots from the posterior to check convergence. The right panels report the corresponding posterior distribution estimate (black solid line) together with the prior distribution for that parameter (red solid line). The \% overlap reported in red is the PPO (prior-posterior overlap).}
    \label{fig:MCMCmulti_line5}
\end{figure}

\end{document}